\documentclass[final,3p,times,twocolumn,authoryear]{elsarticle} 

\usepackage[utf8]{inputenc}
\usepackage{enumitem}
\usepackage{lineno} 
\usepackage{amssymb}
\usepackage{amsmath,bm}
\usepackage{graphicx}
\usepackage{subfig}
\usepackage{multirow}
\usepackage[normalem]{ulem}
\usepackage{appendix}
\usepackage{changepage,geometry}  
\usepackage{multicol}
\usepackage{natbib} 
\usepackage[dvipsnames]{xcolor}
\usepackage{hyperref} 
\hypersetup{hidelinks} 

\usepackage{lineno}

\begin{document}

\begin{frontmatter}



\title{Angular Momentum Drain: Despinning Embedded Planetesimals} 


\author[a1]{Stephen Li}\corref{cor1}
\ead{sluniews@ur.rochester.edu}
\author[a1]{Maggie Ju}
\ead{mju2@u.rochester.edu}
\author[a1]{A. C. Quillen}
\ead{alice.quillen@rochester.edu}
\author[a1]{Adam E. Rubinstein}
\ead{arubinst@ur.rochester.edu} 

\address[a1]{Department of Physics and Astronomy, University of Rochester, Rochester, NY 14627, USA}
 \cortext[cor1]{Corresponding author}
\begin{abstract}

Young and forming planetesimals experience impacts from particles present in a protostellar disk. Using crater scaling laws, we integrate ejecta distributions for oblique impacts. For impacts at 10 to 65 m/s, expected for impacts associated with a disk wind, we estimate the erosion rate and torque exerted on the planetesimal. We find that the mechanism for angular momentum drain proposed by Dobrovolskis and Burns (1984) for asteroids could operate in the low velocity regime of a disk wind. 
Though spin-down associated with impacts can facilitate planetesimal collapse, we find that the process is inefficient. We find that angular momentum drain via impacts operates in the gravitational focusing regime. The angular momentum transfer is most effective when the wind speed is high, the projectile density is high compared to the bulk planetesimal density, and the planetesimal is sufficiently weak that cratering is in the gravity regime. We find that angular momentum drain due to impacts within a pebble cloud may, in those specific conditions, facilitate collapse of single planetesimals.
\end{abstract}




\end{frontmatter}



\section{Introduction}

The spin rates of most asteroids have been altered by impacts \citep{farinella_asteroid_1992,Mao_2020}. 
Because high velocity ejecta preferentially escape along the direction of rotation, impacts cumulatively cause asteroids and other minor bodies to spin down  \citep{DOBROVOLSKIS1984464,cellino_asteroid_1990,takeda_mass_2009,Sevecek_2019}. 
Additional mechanisms for minor body spin evolution include sublimation and outgassing \citep{Mueller_1996,Mottola_2020}, the YORP effect \citep{Bottke_2006} and tidal encounters \citep{scheeres_evolution_2004}. 
In this study we focus on the mechanism for angular momentum drain by impacts proposed by 
\citet{DOBROVOLSKIS1984464}, but apply it in the context of recently formed planetesimals during the epoch when they are embedded in a circumstellar disk. 

Planetesimals are thought to form in particle-rich clumps that coalesce within a protoplanetary disk during an era of streaming instability  \citep{Youdin2005,Johansen_2007,Li_2019,Carrera_2021} or within turbulent vortices \citep{Cuzzi_2008}.  
For variations or alternatives to these scenarios, see the review by \citet{Simon_2022}.
Planetesimal collapse within a cloud of pebbles is expected to incorporate a fraction (between 50\% to 100\%) of a pebble cloud mass into bound objects \citep{Robinson_2020,Nesvorny_2021,Polak_2023}.
During the final states of planetesimal collapse, the angular momentum 
of the pebble cloud can generate a rapidly spinning disk or a bound binary \citep{Nesvorny_2019,Robinson_2020,Nesvorny_2021,Polak_2023}.
A newly formed planetesimal can experience accretion, erosion, and transport of particles across its surface while embedded in the protoplanetary disk 
\citep{Rozner_2020,Demirci_2020b,Quillen_2024}.

\citet{Nesvorny_2021} find that most clumps originate with $l \sim 0.1$ at 45 AU, where $l$ is the scaled angular momentum, $l = L_c / L_H$, $L_c$ is the clump angular momentum, and $L_H$ is the Hill angular momentum.  If a self-gravitating pebble cloud has some spin angular momentum, during collapse it would spin up (e.g., \citealt{Nesvorny_2021}). As these clumps collapse, the angular momentum must be less than the angular momentum of a critically rotating Jacobi ellipsoid, $L^*_J = 0.39\sqrt{GM^3R}$. \citet{Nesvorny_2021} convert this into Hill units, $l^*_J = L^*_J/L_H \sim 0.01$ at 45 AU. This requires angular momentum to be shed for the object to collapse into a single bound object. 

In the context of the formation of Transneptunian binaries, mechanisms for removing angular momentum include the ejection of a third body \citep{2008_Noll},  gravitational interactions with a distribution of small bodies residing in the disk  \citep{Goldreich_2002,Funato_2004}, collisions  \citep{Weidenschilling_2002}, or nebular drag \citep{McKinnon_2020, Lyra2021}.  The simulations by \citet{Nesvorny_2021} illustrate that the particles within the pebble cloud itself can serve as a reservoir for angular momentum, and that collisions between particles within the cloud can facilitate dissipation and angular momentum transfer. 
Binaries, contact binaries, and disk shaped objects may be a consequence of the angular momentum of a pebble cloud prior to its collapse
\citep{Nesvorny_2021,Polak_2023}. 

We focus on the connection between impacts from particles in a pebble cloud and the spin angular momentum of a collapsing body. 
As impacts can spin down an asteroid \citep{DOBROVOLSKIS1984464}, impacts could aid in collapse of a pebble cloud, and the host pebble cloud would serve as a reservoir or sink for angular momentum. While planetesimals are embedded in a disk, low velocity impacts could be frequent \citep{Quillen_2024}. 
While impacts in the asteroid belt tend to be hypervelocity at a few km/s \citep{1994IcarBottke} and can generate a spray of ejected particles \citep{Jewitt_2015,lauretta_episodes_2019}, impact velocities on a planetesimal within a pebble cloud would be substantially lower, between a few to tens of m/s \citep{Johansen_2007,McKinnon_2020,Quillen_2024}. 

With a high velocity ($\sim$ km/s) impact on a small body, most ejecta escapes. For nearly catastrophic impacts, the total escaping ejecta mass is sensitive to pre-impact body rotation \citep{LEINHARDT_2000,BALLOUZ_2015,Sevecek_2019}, impact angle, impact and escape velocities \citep{Hyodo_2021}. For lower velocity impacts,  due to the larger ratio between impact and escape velocities, we expect that the fraction of escaping ejecta would be more sensitive to impact angle and body rotation.  At a few times escape velocity of the larger body or below, the fraction of ejecta that escapes is sensitive to impact velocity \citep{KORYCANSKY_2009,Quillen_2024}, impact angle and body rotation \citep{Ballouz_2014}. 
The works by \citet{DOBROVOLSKIS1984464,Sevecek_2019} focused on high-velocity, erosional impacts that reduce the spin-rate of an asteroid.
We extend this work to a low velocity regime to see how a distribution of impacts affect the mass loss and angular momentum of a planetesimal that is embedded in a protostellar disk. 

The integrations by \citet{DOBROVOLSKIS1984464} neglected the sensitivity of ejecta mass to azimuthal angle for oblique impacts, however recent studies  \citep{raducan_ejecta_2022,Luo_2022,Hirata_2021,Quillen_2024b} have begun to characterize the azimuthal dependence of oblique impact ejecta distributions. In this work,  
we use recent estimates for the azimuthal sensitivity of ejecta to impact angle to determine which ejecta escape. \citet{DOBROVOLSKIS1984464} focused on the role of impacts on asteroids. In this setting the projectiles originate from different directions.  The changes in angular momentum caused by impacts were modeled with a random walk \citep{DOBROVOLSKIS1984464}.   In contrast, the spin axis of a planetesimal that is embedded within a circumstellar disk could be related to the direction of a headwind containing particle projectiles \citep{Nesvorny_2021,Quillen_2024}. Consequently in this study we consider impacts from a uniform direction that is perpendicular to the planetesimal spin axis. 

To form planetesimals within a protostellar disk, dust particles and ices coagulate to form low density and low-strength aggregates, which are often called pebbles \citep{Lorek2018, ORourke2020}. Millimeter sized aggregates, that later form into planetesimals, initially would be  porous, with volume filling factor well below a percent \citet{Lorek2018}.   
An aggregate growing via cohesive or sticky collisions reaches a size known as the bouncing barrier, where collisions bounce rather than stick, compressing the aggregate to a volume filling factor of 1 - 10\% \citep{Lorek2018}. Icy and dusty aggregates are likely porous or fluffy when the disk undergoes streaming instability \citep{Lorek2018}. Because of the large range of possible densities for pebbles in a primordial disk, in this study we also vary the ratio of projectile to planetesimal density.

\subsection{Outline}

In Section \ref{Sec:ImpactsFrames} we discuss reference frames and planetesimal collisions with particles that originate in a headwind. We use the point-source approximation ejecta scaling model developed for normal impacts \citep{Housen_2011} that has been extended to be sensitive to impact angle as well as ejecta azimuthal angle  \citep{raducan_ejecta_2022,Quillen_2024b}. We describe how we estimate the angular momentum transfer and fraction of ejecta that escape from single impacts.

In Section \ref{Sec:SingleCrater} we apply the framework for single impacts to a distribution of collisions across the planetesimal surface. These collisions are oriented on a Cartesian grid overlaid onto the cross-sectional area of the planetesimal in the inertial reference frame. We use the Transneptunian object Arrokoth as a fiducial setting for our integrations and highlight the importance of gravitational focusing. 

In Section \ref{Sec:ParamSweeps}, we vary planetesimal and projectile parameters to explore the sensitivity of the angular momentum transfer efficiency and mass loss ratio to physical properties such as planetesimal and projectile density, planetesimal strength, planetesimal spin rate, and projectile velocity, and spatial projectile velocity gradient. 

\section{Impacts on an Embedded Planetesimal}
\label{Sec:ImpactsFrames}

We consider a planetesimal that is bombarded by particles that reside in a circumstellar disk. 
We assume that, due to a difference between the velocity of the planetesimal and that of gas and small particles in the disk, impacts originate from a preferred direction, as shown in Figure \ref{fig:InertialCoordinates}. 
A planetesimal embedded in the disk feels a wind known as a ‘headwind’ that arises because the gas in the disk is subject to a large-scale pressure gradient, giving the gas a mean tangential circular velocity below that of a circular Keplerian orbit \citep{Whipple_1972}.
We refer to the particle flow as particles within a headwind, but impacts from particles could also occur within a vortex clump \citep{Cuzzi_2008} or a pebble cloud clump that has formed during streaming instability \citep{Johansen_2007}.

\subsection{Planetesimal Properties}

Following \citet{DOBROVOLSKIS1984464}, 
we take the planetesimal to be spherical with uniform escape velocity, 
\begin{equation}
u_{esc} = \sqrt{2GM_a/R_a} \label{Eq:EscVel}
\end{equation}
where $M_a$ is the planetesimal mass, $R_a$ is the planetesimal radius, and $G$ is the gravitational constant. 
We assume solid body rotation and an isotropic and homogeneous body.

Because planetesimals can have a high spin rate \citep{Johansen_2010,Robinson_2020,Nesvorny_2021}, 
we describe body spin in terms of its maximum rotation rate where a particle on the equator would leave the surface due to centrifugal acceleration.   We put spin in units of the centrifugal break-up value for a sphere, 
\begin{align}
    \Omega_{\rm break-up} &=\sqrt{ \frac{4\pi G \rho_a}{3}},  \label{Eq:MaxRot}
\end{align}
where $\rho_a$ is the planetesimal's bulk density.

\begin{figure}[htbp]
    \centering
    \includegraphics[width = 2.25truein]{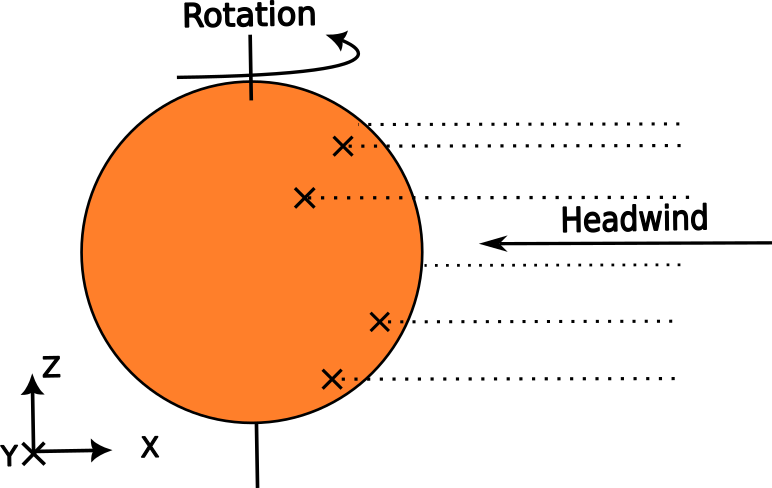}
    \caption{An image indicating the inertial reference frame coordinate system. The direction of planetesimal rotation is shown with respect to the headwind. Dashed lines show trajectories of headwind particles that collide with the planetesimal.  We assume that the planetesimal spin axis is perpendicular to the circumstellar disk plane. The star is located in the positive y direction.}
    \label{fig:InertialCoordinates} 
\end{figure}

The rotation axis of the planetesimal is assumed to be perpendicular to the incident
wind direction, as shown in Figure \ref{fig:InertialCoordinates}.
This assumption is motivated by simulations of pebble accretion which suggest that planetesimals form with rotation axis prograde with respect to the disk in which they are embedded \citep{Johansen_2010}.  
Figure \ref{fig:InertialCoordinates} illustrates a Cartesian inertial coordinate system, with wind particles originating from the $+x$ direction and the planetesimal spin axis pointing toward the $+ z$ direction. 

\subsection{Inertial and Rotation Frames}
\begin{figure}[htbp]\centering
\includegraphics[width=7cm]{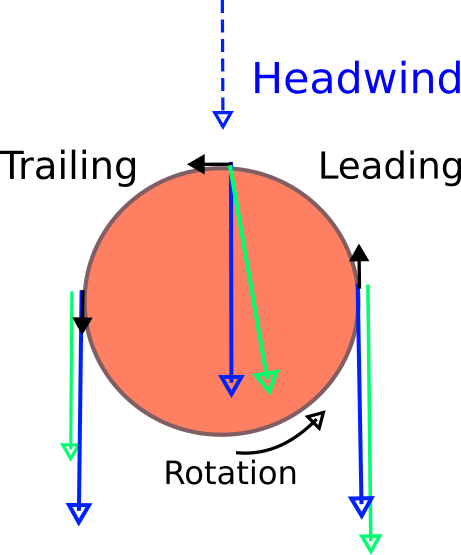}
\caption{We illustrate how rotation affects impact angle and velocity in as seen in frame rotating with the surface. Black shows rotation.  Blue shows wind speed. Green shows wind particle velocity in the frame rotating with the planetesimal.  From the perspective of a viewer on the surface on the leading side, rotation causes the wind particle to impact the surface at higher velocity and at lower (closer to grazing) impact angle. }
\label{fig:wind_on_sphere} 
\end{figure}

The impact velocity for a collision with a headwind particle is 10 - 65 m/s (e.g., \citealt {Quillen_2024}) and is low compared to most impacts that currently take place in the inner solar system, which are at velocities of a few km/s \citep{1994IcarBottke}. A 10 km planetesimal of bulk density $\rho_a = 1000$ kg m$^{-3}$ spinning near breakup has rotational velocity of about 5 m/s. 
The ratio of rotational velocity to impact velocity and the ratio of rotational velocity to the velocity of ejecta are higher in our setting than in the context of km/s impacts on asteroids (studied by \citealt{DOBROVOLSKIS1984464}).  
As a consequence we do not neglect the rotational velocity when computing impact velocities in the frame rotating with the planetesimal surface. 
Figure \ref{fig:wind_on_sphere} illustrates how the impact velocity, as seen in a frame rotating with the surface, depends upon the impact location and the direction of rotation.

We denote the location of an impact on the planetesimal surface with the vector  ${\bf r}$, originating from the planetesimal's center of mass. The velocity of the surface at ${\bf r}$ is 
\begin{equation}    
    {\bf u}_{rot} = {\boldsymbol \Omega}_a \times {\bf r} .
    \label{eq:vrot}
\end{equation} 
where ${\boldsymbol \Omega}_a$ is the planetesimal spin vector, which we take to be aligned with $\hat{z}$. On a spherical planetesimal ${\bf r} = R_a \hat {\bf n}.$

\subsection{The Impact Velocity and Impact Angle in the Rotating Frame}

\begin{figure*}[htbp]
    \centering
    \includegraphics[width = 10cm]{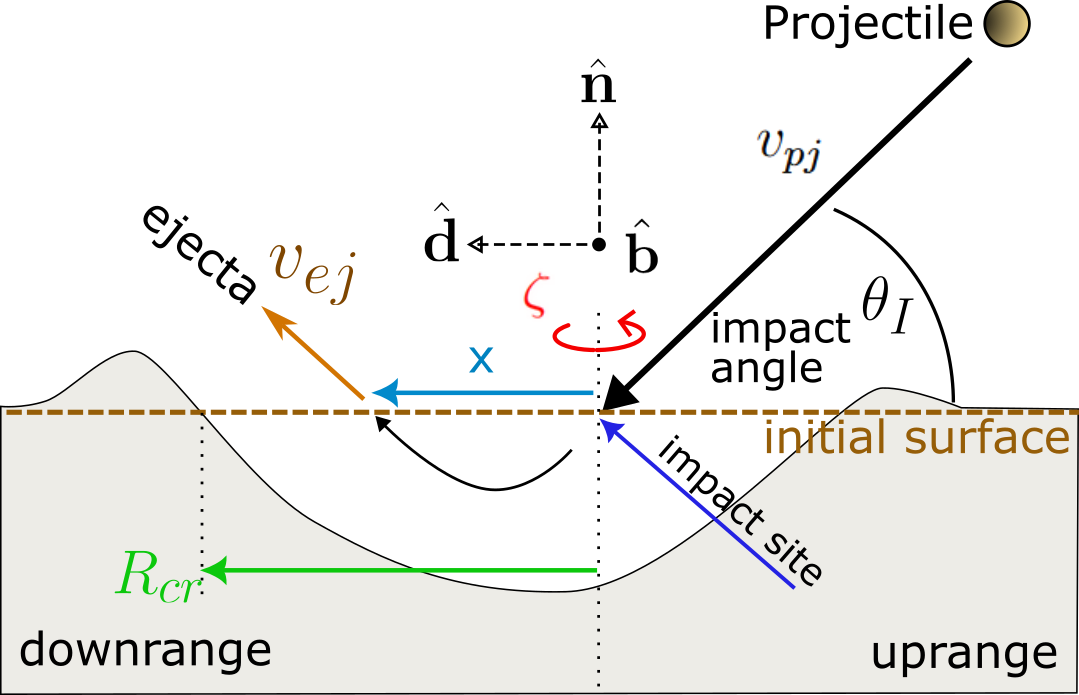}
    \caption{In the frame rotating with the surface, we illustrate variables relevant for describing crater, projectile, and ejecta.  The variables  $x, \zeta$ are coordinates measured with origin at the site of impact. The normal unit vector $\hat{\bf{n}}$, the down-range unit vector $ \hat{\bf{d}}$, and $\hat{\bf{b}}$ are orthogonal vectors defining a local coordinate system. The crater radius $R_{cr}$ is the radius from the site of impact where the crater profile crosses the surface level (before impact) and it can be a function of impact angle $\theta_{imp}$ and azimuthal angle $\zeta$. The ejecta velocity is $v_{ej}$.}
    \label{fig:CraterUnits} 
\end{figure*}

The ejecta velocity distribution depends on both impact velocity and direction in the frame rotating with the surface at the point of impact.
In the rotating frame moving with the surface, the projectile velocity is
\begin{align}
{\bf v}_{pj} = {\bf u}_{pj} - {\bf u}_{rot} \label{eqn:vrotpj}
\end{align}
where ${\bf u}_{pj}$ is the projectile velocity in the inertial frame, as shown in Figure \ref{fig:InertialCoordinates}. The rotation velocity ${\bf u}_{rot}$ is calculated using Equation \ref{eq:vrot} for a given position vector ${\bf r}$. 

We define a unit vector $\hat {\bf n}$ as vector normal to the surface and pointing away from the surface at the site of impact.  We define a downrange vector $\hat{\bf d}$ as a unit vector tangential to the surface pointing downrange (or away) from the direction from which the projectile originated.  
The unit normal and downrange vectors are 
\begin{align}
\hat{\bf{n}} &= \frac{ {\bf r} }{r} 
\label{Eq:nvec}\\
\hat {\bf d} &=  \frac{ 
{\bf v}_{pj} - ( {\bf v}_{pj} \cdot \hat {\bf n} )\hat {\bf n} } 
{| 
{\bf v}_{pj} - ( {\bf v}_{pj} \cdot \hat {\bf n} )\hat {\bf n} |} 
\label{Eq:dvec}
\end{align}
where ${\bf v}_{pj}$ is the projectile velocity vector 
in the frame rotating with the surface. The unit vector $\hat {\bf d}$ is computed by projecting the incoming projectile velocity into the plane tangent to the surface at the point of impact. 
It is convenient to have a set of 3 orthogonal unit vectors at the site of impact.  A third unit vector, $\hat {\bf b}$,  called the binormal vector, is perpendicular to $\hat {\bf n}$ and $\hat {\bf d}$, and 
is 
\begin{equation}
    \hat{{\bf b}} = \hat{{\bf n}} \times \hat{{\bf d}}.
    \label{Eq:bvec}
\end{equation} 

The impact angle, $\theta_{imp}$ is defined as the angle between the projectile velocity (in the rotating frame) and the plane tangent to the surface 
\begin{equation}
    \theta_{imp} = \arcsin{\left(\frac{-{\bf v}_{pj} \cdot \hat {\bf n} }{|{\bf v}_{pj}|}\right)} .\label{eqn:theta_imp}
\end{equation}
The minus sign arises because we have defined the normal direction $\hat {\bf n}$ pointing outward from the surface.  The impact angle is low for grazing impacts and equal to $\pi/2$ for a normal impact. Figure \ref{fig:CraterUnits} illustrates 
the different directions and angles for an impact in the rotating frame.

\subsection{Crater Scaling and Ejecta Distributions}
\label{subsec:cs}

The momentum transfer parameter and the angular momentum imparted to a planetesimal by an impact depends upon the integrated properties of the ejecta (e.g., \citealt{DOBROVOLSKIS1984464,takeda_mass_2009,Hyodo_2021,raducan_ejecta_2022}). We first discuss the velocity and mass distributions of ejecta from normal impacts and then consider more general ejecta distributions from oblique impacts.

When computing the properties of ejecta, we assume that planetesimal mass exceeds projectile mass, $M_a \gg m_{pj}$,  and planetesimal mass also exceeds total ejecta mass; $M_a \gg M_{ej}$. We neglect the possibility of projectile ricochet (e.g., \citealt{Wright_2020}) and projectile spin (e.g., \citealt{2023aCarvalho,2023bCarvalho,2024Carvalho}). 

Normal impacts are often described in terms of three dimensionless parameters \citep{Housen_2011}; 
\begin{align}
\pi_2 & = \frac{g_a a_{pj} }{  v_{pj}^2 } \label{Eq:pi2} \\
\pi_3 &= \frac{Y_a}{\rho_a v_{pj}^2} \label{Eq:pi3} \\
\pi_4 &=  \frac{\rho_a}{\rho_{pj}} \label{Eq:pi4}
\end{align}
where $g_a$ is the surface gravitational acceleration on the planetesimal surface, the projectile radius is $a_{pj}$, $Y_a$ is the strength of the planetesimal material, and the density of planetesimal and projectile are $\rho_a$ and $\rho_{pj}$, respectively.

The material properties we assume for strength regime craters is sand/fly ash (SFA), and for gravity regime craters is sand. These materials have been used as material analogues in order to study impact cratering into porous asteroids \citep{Housen2003,Housen_2011, Housen_2018}.
We adopt the ejecta scaling laws for normal impacts by \citet{Housen_2011}, which were derived assuming a point source approximation. While scaling laws are often developed in a higher velocity regime, we are applying these laws in a  low-velocity regime as the experimental data compiled by \cite{Housen_2011} for ejecta scaling of normal impacts cover a wide range in impact speed, including impacts as low as 6 m/s.

The ejection velocity is a function of distance $x$ from the impact site where the ejecta crosses the surface level plane (as it was prior to impact), as illustrated in Figure \ref{fig:CraterUnits}.  
The ejecta velocity scaling law (Eq 14 from \cite{Housen_2011}) is
\begin{align}
    \frac{ v_{ej,N}(x, \pi_4)}{v_{pj}} & =  C_{1,N} 
    \left(\frac{x}{a_{pj}}\right)^{-\frac{1}{\mu_N}}
    \pi_4^{-\frac{\nu}{\mu_N}}
    \left(1- \frac{x}{n_2 R_{cr,N}} \right)^p . 
   \label{Eq:EjVel}
\end{align}
The exponents $\mu_N, \nu$ and coefficients $C_{1,N}$ depend 
upon substrate material properties. 
We have added an $N$ subscript, denoting normal impact, to parameters that we vary below when discussing oblique impacts. 

\citet{Housen_2011} find that the 
coefficients $n_1 \sim 1.2$,   $n_2 \sim 1$, and $p \sim 0.3$. 
Equation \ref{Eq:EjVel} holds for $ n_1 a_{pj}< x < n_2 R_{cr,N}$, where
 $R_{cr,N}$ is the crater radius for a normal impact.  The crater radius
 is the radius at which the crater profile crosses the surface level plane, as shown in Figure \ref{fig:CraterUnits}. 
In the strength regime, the crater radius for a normal impact is
\begin{align}
   R_{cr,N}(\pi_3,\pi_4) = a_{pj} \left(\frac{4\pi}{3} \right)^\frac{1}{3} H_2 
\pi_3^{-\frac{\mu_N}{2}} \pi_4^{(1-3\nu)/3}, \label{Eq:Rcr_str}
\end{align}
with coefficient $H_2 \sim 0.4$ that also depends upon substrate material properties \citep{Housen_2011}. In the gravity regime, the crater radius for a normal impact is
\begin{equation}
    R_{cr,N}(\pi_2,\pi_4) = a_{pj}\left(\frac{4\pi}{3}\right)^{\frac{1}{3}}H_1\pi_2^{\frac{-\mu_N}{2+\mu_N}}\pi_4^{\frac{2+\mu_n-6\nu}{6+3\mu_N}}, \label{Eq:Rcr_grav}
\end{equation}
with coefficient $H_1 \sim 0.59$. Crater scaling parameters and materials used for both strength and gravity regimes can be found in Table \ref{tab:craterScaling}.

The ejecta mass integrated between $n_1 a_{pj}< x < n_2 R_{cr,N}$  from the site of impact is
\begin{align}
    \frac{M(<x, \pi_4)}{m_{pj}} = \frac{3k_{N}}{4\pi} \pi_4 
    \left[\left( \frac{x}{a_{pj}} \right)^3 - n_1^3 \right] \label{eqn:EjMass}
\end{align}
and with coefficient $ k_{N} \sim 0.3$ \citep{Housen_2011}.  Section \ref{sec:regime}, below, contains additional discussion on the impact regime. 

\subsubsection{Ejecta distributions for oblique impacts}

Ejecta velocity, ejecta angle, and mass distributions for oblique impacts are sensitive to both impact angle, $\theta_{imp}$, and azimuthal angle within the crater, $\zeta$ \citep{Anderson_2003,raducan_ejecta_2022,Quillen_2024b}. We take azimuthal angle $\zeta$ to be the angle in the surface plane (as seen from above the crater in the plane spanned by unit vectors $\hat {\bf d}$ and $\hat {\bf b}$), with direction $\zeta = 180^\circ$ downrange, as adopted by \citet{Anderson_2003,raducan_ejecta_2022}.

Following \citet{raducan_ejecta_2022,Quillen_2024b} we modify equations \ref{Eq:EjVel}, \ref{Eq:Rcr_str} or \ref{Eq:Rcr_grav},  and \ref{eqn:EjMass} to be sensitive to both impact angle and azimuthal angle, giving  
functions for crater radius $R_{cr}$,  exponent $\mu$, and coefficients $C_1, k$ (dropping the $N$ subscript) 
that are consistent with normal impact scaling 
\begin{align}
    \mu( \zeta, \theta_{imp} = \pi/2) &= \mu_N \\
    C_1 (\zeta, \theta_{imp} = \pi/2) &= C_{1,N} \\
    R_{cr}(\zeta,\theta_{imp}=\pi/2) & = R_{cr,N} \label{Eq: craterRadius} \\
    \int_0^{2 \pi} k(\zeta,\theta_{imp}=\pi/2) d \zeta &= k_N . 
\end{align}
We adopt the functions estimated by \citet{raducan_ejecta_2022} from measurements of ejecta from 3D iSALE oblique impact simulations; 
\begin{align}
    \mu(\zeta,\theta_{imp}) &= \mu_N\left( 1  + \frac{1}{2}\cos \zeta \cos \theta_{imp}  \right)  \label{Eq: mu}\\
    C_1(\zeta,\theta_{imp}) & = C_{1,N} \exp (-5 \cos \zeta \cos \theta_{imp}) \label{Eq: C1} \\
    R_{cr}(\zeta,\theta_{imp}) & = R_{cr,N} \left[ 1 - \frac{9 - \frac{18}{\pi}\theta_{imp}}{10} \frac{\cos \zeta}{2}\right] 
    \label{eq:raducanRadius} \\
    k(\zeta,\theta_{imp})  & = \frac{k_N}{2\pi}
    \exp (-0.02 \cos \zeta \cos \theta_{imp}) . \label{Eq: k} 
\end{align}
In equation \ref{eq:raducanRadius} the impact angle $\theta_{imp}$ is in radians.

The functions given by \citet{raducan_ejecta_2022} in their equations 6--9 (and here with our equations \ref{Eq: mu} -- \ref{Eq: k}) are not consistent with their Figure 12 but appear consistent with the distributions shown in their other figures.   The weak dependence of the function $k(\zeta,\theta_{imp})$ on angles $\theta_{imp}$ and $\zeta$ is supported by laboratory measurements of the azimuthal ejecta mass distribution from $\sim100$ m/s oblique impacts into sand \citep{Quillen_2024b}. 

With the scaling functions for oblique impacts (Equations \ref{Eq: mu} - \ref{eq:raducanRadius}), we modify Equation \ref{Eq:EjVel}, giving an oblique impact ejecta velocity function,
\begin{align} \frac{
v_{ej}(x, \pi_4, \theta_{imp},\zeta)}{v_{pj}} = &\ C_1(\theta_{imp},\zeta)  x^{-\frac{1}{\mu(\theta_{imp},\zeta)}} 
 \pi_4^{-\frac{\nu}{\mu(\theta_{imp},\zeta)}} 
 \nonumber \\ & \ \times
\left(1 - \frac{x}{n_{2} R_{cr}(\theta_{imp}, \zeta)} \right)^{p}  , 
\label{Eq:ObliqueEjectaVel}
\end{align}
where $\mu(\theta_{imp},\zeta), C_1(\theta_{imp},\zeta)$ and $R_{cr}(\theta_{imp},\zeta)$ are calculated from Equations \ref{Eq: mu} --  \ref{eq:raducanRadius}. Equation \ref{eqn:EjMass} for the mass distribution, is similarly modified to depend on azimuthal and impact angles  
\begin{align}
    \frac{M(<x, \pi_4, \theta_{imp}, \zeta)}{m_{pj}} = \frac{3k(\theta_{imp}, \zeta)}{4\pi} \pi_4 
    \left[\left( \frac{x}{a_{pj}} \right)^3 - n_1^3 \right] \label{eqn:AngularEjMass}
\end{align}
with $k(\theta_{imp}, \zeta) $ from equation \ref{Eq: k}.

\subsection{Ejecta Velocity Components}

\begin{figure}[htbp]
    \centering
    \includegraphics[width = 8cm]{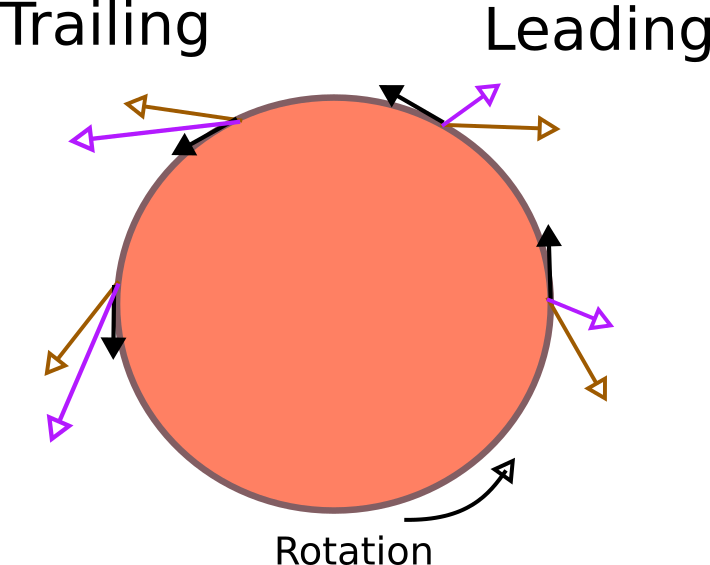}
    \caption{We illustrate how rotation affects the ejecta velocity as seen in an inertial frame.   Black arrows show rotation, Brown shows ejecta particle velocity in the frame rotating with the surface at an angle from normal of about $45^\circ$. To transfer into the inertial frame, we add the rotational velocity, resulting in the purple arrows showing the ejecta velocity in the inertial frame.  The purple arrows are longer on the trailing side, implying that angular momentum is carried away by ejecta. Ejecta curtains are computed in the frame rotating with the surface (brown arrows), however the likelihood of escape depends on whether the ejecta velocity  in an inertial frame (purple arrows), exceeds the escape velocity.}
    \label{fig:EjectaFrameTransfer} 
\end{figure}

Ejecta distributions are calculated in the frame rotating with the surface; however, to determine which ejecta escape the planetesimal, we compare the ejecta velocities in the inertial frame to the escape velocity.    
In the surface rotational frame, and 
in terms of the surface normal, downrange and binormal vector directions, 
the ejecta velocity is 
\begin{align}
    {\bf v}_{ej}  = v_{ej,d} \hat {\bf d} + v_{ej,b} \hat {\bf b}
    + v_{ej,n} \hat {\bf n}\label{Eq:vejn}
\end{align}
with components
\begin{align}
    v_{ej,d}   &= v_{ej} \cos(\zeta + \pi) \cos(\theta_{ej}) \nonumber \\
    v_{ej,b}   &= v_{ej} \sin(\zeta + \pi) \cos(\theta_{ej}) \nonumber \\
    v_{ej,n}   &= v_{ej} \sin(\theta_{ej}).  \label{Eq:vejn_components}
\end{align}
Here $v_{ej,d}$ is the downrange component, $v_{ej,n}$ is the normal component, $v_{ej,b}$ is the binormal component, and $\theta_{ej}$ is the ejecta angle.  
Experiments and simulations find that 
the ejecta angle $\theta_{ej}$ varies by less than 10 degrees as a function of impact angle, launch position and azimuthal angle \citep{Anderson_2003, raducan_ejecta_2022}.   Hence we assume that 
$\theta_{ej}$ is constant. 


For each impact we compute 
 ejecta distributions  
in the frame rotating with the surface  using equations \ref{Eq:ObliqueEjectaVel} for the velocity and \ref{eqn:AngularEjMass} for the mass distribution. To determine which ejecta escape the planetesimal, we transfer the ejecta velocity back into the inertial frame by reversing equation \ref{eqn:vrotpj};
\begin{align}
{\bf u}_{ej}  = {\bf v}_{ej} + {\bf u}_{rot}. \label{eqn:inertialConversion}
\end{align}

\subsection{The Mass and Momentum of Escaping Ejecta}
Following \citet{Holsapple_2012}, we assume that 
a parcel of ejecta with velocity greater than the escape velocity,  $ u_{ej} \ge u_{esc}$,   escapes from the system.  We define a function that describes whether a parcel of ejecta escapes,
\begin{equation}
    V(x,\pi_4,\theta_{imp},\zeta) = \begin{cases} 
          1, & u_{ej}(x,\pi_4,\theta_{imp},\zeta) \ge u_{esc} \\
          0, & u_{ej}(x,\pi_4,\theta_{imp},\zeta) < u_{esc} 
       \end{cases}, 
\end{equation}
where the escape velocity is from Equation \ref{Eq:EscVel}. 

To find the ejecta mass for a specific launch location $x$ from the site of impact, we differentiate Equation \ref{eqn:EjMass} (with respect to $x$) giving 
\begin{equation}
\frac{ d\left( M(<x,\pi_4,\theta_{imp},\zeta)\right)}{m_{pj}} = \frac{9k(\theta_{imp},\zeta)}{4\pi} \pi_4 \left(\frac{x^2}{a_{pj}^3}\right) dx d\zeta.
\label{eqn:dEjMass}    
\end{equation}
Integrating this over possible launch position $x$ within the crater radius and for different azimuthal angles $\zeta$, we compute the total escaping ejecta mass from a single impact; 
\begin{equation}
M_{ej,esc} = \int_x \int_{\zeta}d\left(M(<x,\pi_4,\theta_{imp},\zeta)\right) V(x,\pi_4,\theta_{imp},\zeta) .\label{Eq:MassIntegration}
\end{equation}
Since we neglect ricochet, we assume the mass of the projectile is either accreted onto the planetesimal or incorporated into ejecta.  Thus $M_{ej,esc}/m_{pj} < 1$ indicates accretion, while $M_{ej,esc}/m_{pj} > 1$ indicates erosion.

For a distribution of impacts, we define a mass ratio 
\begin{align}
    \epsilon_M \equiv  \biggl< \frac{M_{ej,esc}}{m_{pj}}  \biggl>
\end{align}
which is the average fraction of mass ejected per unit projectile mass. 

We denote the momentum in escaping ejecta in the rotating frame as ${\bf p}_{ej,esc}$. 
We calculate the downrange component of the escaping ejecta 
\begin{align}
{\bf p}_{ej,esc} \cdot \hat {\bf d} = &  \int_x \int_{\zeta} ({\bf v}_{ej} \cdot \hat {\bf d}) \ 
d\left(M(<x,\pi_4,\theta_{imp},\zeta)\right) 
\times \nonumber \\ & \qquad \ \ 
V(x,\pi_4,\theta_{imp},\zeta) 
\label{Eq:LinearMomentum}.
\end{align}
The $\hat {\bf b}$ momentum component of the escaping ejecta can be neglected because the adopted ejecta distribution is symmetric about the $\hat {\bf d}$-$ \hat{\bf n}$ plane. 
We don't compute the normal momentum component because it does not contribute to the torque on the planetesimal.

\subsection{The Angular Momentum Transfer Efficiency}
\label{sec:eff}

The change in planetesimal angular momentum due to the impact at location ${\bf r}$  would be 
\begin{align}
\Delta {\bf L}_{a} = m_{pj}  {\bf r} \times {\bf u}_{pj} - {\bf r} \times {\bf p}_{ej,esc} \label{eqn:DeltaLa}
\end{align}
repeating equation \ref{eqn:Delta_La}.
See \ref{app:Momentum} for why this depends upon the momentum of escaping ejecta computed in the rotating frame and why we can neglect the ejecta trajectory after impact. 

When averaging over a distribution of impacts that lack angular momentum with respect to the planetesimal, the projectile momentum term in equation \ref{eqn:DeltaLa} can be neglected as it averages to zero. We only need to calculate the rightmost term that depends on the momentum of escaping ejecta as seen in the rotating frame. 
Because the planetesimal is orbiting the star, it is likely to feel a headwind coming from different directions in the orbital plane (see Figure \ref{fig:InertialCoordinates}). We assume that $\Delta {\bf L}_a$ components that are not aligned with the planetesimal's spin axis average to zero. Within a crater, the crater ejecta is symmetric about the uprange-downrange line, and thus the binormal components average to zero. The normal component of the crater ejecta do not average to zero due to asymmetry in the projectile impact velocity in the rotating frame. This would lead to variations in the radial position of the planetesimal orbit and is a consideration for future work. The spin angular momentum of the planetesimal would not be affected by this, but rather the orbital angular momentum. This implies that the average change in spin angular momentum per impact for a distribution of projectiles is 
\begin{align}
\langle \Delta {\bf L}_a \rangle = \left( \bigl<
 {\bf r} \times ({\bf p}_{ej,esc} \cdot \hat{\bf{d}})\hat{\bf{d}} \bigl>  \ \cdot  \ 
     \hat {\boldsymbol \Omega}_a \right)  
    \hat {\boldsymbol \Omega}_a
     \label{eqn:ddl2}
\end{align}
where 
$\hat {\boldsymbol \Omega}_a$ is the unit vector along the spin axis. 
We only need to use the downrange component of escaping ejecta (equation \ref{Eq:LinearMomentum}) to calculate the angular momentum parallel to the spin axis, as that is transferred to the planetesimal via the escaping ejecta. 
The change in angular momentum of the planetesimal 
per unit projectile mass is equal to $\langle  \Delta {\bf L}_a\rangle/m_{pj} $. 

We define a dimensionless angular momentum transfer efficiency 
\begin{align}
    \epsilon_L \equiv - \frac{|\langle \Delta {\bf L}_{a} \rangle|}{L_a}  \frac{M_a}{m_{pj}}
    \label{eqn:effiency}
\end{align}
where $L_a = \frac{2}{5} M_a R_a^2 \Omega_a$ is the planetesimal's spin angular momentum and that of a homogeneous spinning sphere.      

The angular momentum transfer efficiency is the fractional change in angular momentum caused by a distribution of projectiles per unit projectile mass in units of planetesimal mass.   The amount of projectile mass required to remove a significant fraction of the planetesimal's spin angular momentum would be approximately $M_a /\epsilon_L$.

\begin{table*}
\centering
\begin{tabular}{llll}
\hline
\multicolumn{3}{c}{Fiducial Model Parameters}\\
\hline 
Parameter  & Symbol  & Fiducial Value  \\
\hline
Impact Regime  &     &  Strength \\
Planetesimal Radius & $R_a$ & 10 000 [m] \\
Planetesimal Density & $\rho_a$ & 235 [kg m$^{-3}$] \\
Planetesimal Strength & Y$_a$ & 100 [Pa] \\ 
Planetesimal Mass & M$_a$ & $\frac{4\pi}{3}\rho_a R_a^3$ [kg]\\
Planetesimal Rotation Velocity & $\Omega_a$& $\Omega_{\rm break-up}$\\
Headwind Velocity (Inertial) & $u_{hw}$ & 30 [m/s] \\
Projectile Density & $\rho_{pj}$ & 235 [kg m$^{-3}$] \\
Ejecta Angle &  $\theta_{ej}$ & 45$^{\circ}$ \\ \hline
\end{tabular}
\caption{Planetesimal fiducial parameters assumed throughout the simulations unless other values are stated. These parameters are based on the Kuiper Belt Object (486958) Arrokoth \citep{McKinnon_2020,keane_geophysical_2022, 2022Brisset}, with the radius being the spherical equivalent from the mass and density of Arrokoth. \citealt{McKinnon_2020} give a minimum cohesive strength range of 100-400 Pa for Arrokoth determined as a minimum required strength for stabilization of the neck that Arrokoth exhibits. The projectile velocity is taken to be a value near the headwind velocity experienced by planetesimals in the outer disk \citep{Quillen_2024}. For these fiducial parameters, the escape velocity from the planetesimal is 3.6 m/s. }
\label{tab:FidParams}
\end{table*} 

\begin{table*}
\centering
\begin{tabular}{cccccccccccc}
\hline 
\multicolumn{12}{c}{Crater Scaling Parameters}\\ \hline
Regime & Material & $C_{1,N}$ & $k_N$ & $\nu$ & $\mu_N$ & $n_{1}$ & $n_2$ & $H_1$ & $H_2$ & $p$ & $q$ \\
\hline
Strength & SFA & 0.55 & 0.3  & 0.4 & 0.4 & 1.2 & 1.0 & - &0.4 & 0.3 & 0.2\\
Gravity & Sand & 0.55 & 0.3  & 0.4 & 0.41 & 1.2 & 1.3 & 0.59 & - & 0.3 & 0.2\\
\hline
\end{tabular}
\caption{Crater scaling coefficients are taken from Table 3 by \citet{Housen_2011} for SFA (sand/fly ash) and for sand. We adopt parameters typical of granular materials.}
\label{tab:craterScaling}
\end{table*} 


\subsection{Gravitational Focusing}
\label{sec:grav_foc}

For higher mass planetesimals, those with escape velocity above the projectile velocity, the gravitational field of the planetesimal causes gravitational focusing.  In this case the velocity and angle of impact is sensitive to the orbit of the projectile when approaching the planetesimal.  

We compute the velocity and location of the impact using a hyperbolic orbit, neglecting gas drag. To collide with a spherical object of mass $M_a$ and radius $R_a$, a projectile must have impact parameter $b$ lower than 
\begin{equation}
b_{max} = R_a \sqrt{1 + \frac{2 GM_a}{u_{hw}^2 R_a}}.
\label{eq:bmax}
\end{equation}
We have neglected the size of the projectile, assuming that it is small compared to the planetesimal. 
Here $u_{hw}$ is the headwind velocity distant from the planetesimal.  We assume that the projectile has the same velocity. 
With an impact parameter of $b_{max}$, the projectile would graze the planetesimal.

We define a new Cartesian coordinate system with coordinates $(x_o, y_o, z_o)$ to describe the hyperbolic orbit of the projectile.  
The orbit of the projectile lies in the $x_o,y_o$ plane.  The angle 
 $\alpha$ is the angle between the projectile orbital plane and the xy plane of the circumstellar disk.
We assume that the projectile arrives from the $x_o = + \infty$ direction. 
We determine the components of the impact location on the planetesimal surface 
\begin{align}
    {\bf r}_{impact} = R_a (\cos (\Delta f) \hat {\bf x}_o + \sin (\Delta f) \hat {\bf y}_o ).
\end{align}
where $\Delta f$ is the difference between the maximum true anomaly and the true anomaly at impact.  
See \ref{app:Hyperbolic}, and specifically equation \ref{eqn:Delta_f}, for an expression for $\Delta f$. 

We calculate the impact velocity in the inertial frame using energy conservation,
\begin{align}
|\textbf{u}_{pj}| = \sqrt{u_{hw}^2 + \frac{2 GM_a}{R_a}}.
\end{align}
Only when we neglect gravitational focusing entirely is the velocity at impact equal to that of the headwind, $u_{pj} = u_{hw}$. 
Even on low mass planetesimals, with $u_{hw} < u_{esc}$, the gravitational field would affect the impact velocity $u_{pj}$. 
The collision angle (the angle between the surface normal and the orbital trajectory at the impact point) is 
\begin{align}
\gamma_c = {\rm atan}\left( \frac{1 + e \cos f_{impact}}{e \sin f_{impact}}\right)
\end{align}
with the convention that a grazing impact has $\gamma_c = \frac{\pi}{2}$. See \ref{app:Hyperbolic} Equation \ref{Eq:fimp} for an equation for the true anomaly of the impact, $f_{impact}$.
The impact velocity vector in the inertial reference frame is then
\begin{align}
{\bf u}_{pj} = -|u_{pj}|\left( \cos (\Delta f - \gamma_c) \hat {\bf x}_o
+ \sin (\Delta f - \gamma_c) \hat {\bf y}_o\right).
\label{eqn:gravfocusvel}
\end{align}
We calculate the projectile velocity using Equation \ref{eqn:gravfocusvel} for all impacts, regardless of the planetesimal mass.
 
To go into the $x,y,z$ coordinate system relevant for computing the ejecta distributions, shown in Figure \ref{fig:InertialCoordinates}, 
(taking into account the orbit orientation) we rotate the calculated location of impact and the velocity of impact around the x-axis with rotation matrix 
\begin{align}
R(\alpha)_x =
\begin{pmatrix} 
1 & 0 & 0 \\
0 & \cos \alpha & -\sin \alpha \\
0 & \sin \alpha & \cos \alpha \\
\end{pmatrix}.
\label{Eq:rotationMat}
\end{align}
The resulting impact velocity is used to compute the impact angle in the rotating frame, with Equation \ref{eqn:theta_imp}. 

\subsection{Caveats}

While computing the integrated erosion rate and the angular momentum transfer efficiency in this section, we have neglected some degrees of freedom. We mention these points here as possible directions of future study.

\subsubsection{Gravitational Focusing}
We have neglected gravitational focusing of the escaping ejecta particles. Neglecting gravitational "defocusing" does not influence our estimate of the angular momentum transfer efficiency, but would influence computation of the linear momentum transfer efficiency. 

\subsubsection{Planetesimal Shape and Structure}

For simplicity, we have neglected the shape of the planetesimal by assuming that it is spherical.  If we allow for non-spherical planetesimals, the gravity vector at a surface location is not necessarily perpendicular to the surface, so the surface can have a slope with respect to the direction of self-gravity.  Studies measuring crater volume, diameter, and axis ratio of oblique impacts into inclined granular media find that ejecta distributions also depend on surface slope \citep{Takizawa_2020, Omura_2021}.  In future, models could be extended to take into account impact angle, surface slope and the direction and values of ejecta velocity vectors that allow ejecta to escape.  Such models could find an analogies to the Yarkovsky effect, causing orbital drift, but mediated by momentum exchange with a distribution of projectiles.  

We neglected heterogeneity of the planetesimal surface. Its surface could be rough and contain particles of different sizes, densities, compositions, and shapes. These variations are not accounted for in the ejecta distribution models that are used throughout this paper but might influence the fraction and momentum content of escaping ejecta.

\subsubsection{Gas Drag Around Planetesimals}
Crater ejecta would be effected by the headwind flowing around the surface of the planetesimal, as shown by \cite{Quillen_2024}. Accurate models of rarefied gas flow around a planetesimal would help determine the quantity of ejecta that escapes, as small ejecta particles are swept up by a disk wind near the planetesimal surface. A wind would accelerate small particles in low velocity ejecta to beyond the escape velocity. However, the fraction of material swept up would be approximately the same equal on leading and trailing sides of the planetesimal, so we suspect that the total contribution to the angular momentum loss would be low.

\subsubsection{Rotating Projectiles}
In this study we have neglected projectile spin. However, studies of impacts from spinning projectiles \citep{2023aCarvalho, 2023bCarvalho, 2024Carvalho} find that crater profiles are sensitive to projectile spin.
To accurately take into account the projectile spin, we would need an accurate description of the spin angular momentum distribution of the particles within the source of projectiles. Additionally, crater scaling coefficients for the ejecta could become dependent upon the projectile spin.

\section{Impact Regime}

\subsection{Gravity vs Strength Impact regime}
\label{sec:regime}
Our numerical procedure for calculating the angular momentum transfer efficiency $\epsilon_L$ and fraction of mass per unit projectile mass  $\epsilon_M$ depend on whether impacts are in the gravitational or strength regime through the crater radius on the regime (equations \ref{Eq:Rcr_str} and \ref{Eq:Rcr_grav}). 
The transition between gravity regime and strength regimes depends on the dimensionless $\pi_2, \pi_3, \pi_4$ parameters \citep{Holsapple1993}.  The gravity regime is where  
\begin{equation}
    \pi_2 \gtrsim \left(\frac{H_1}{H_2}\right)^{(2+\mu_N)/\mu_N}\pi_4^{\nu}\pi_3^{(2+\mu_N)/2}
\end{equation}
(following \citealt{Holsapple1993,Housen_2011}).
We solve for the projectile velocity at the transition between regimes, 
\begin{align}
v_{pj}\gtrsim \left(\frac{Y_a}{\rho_a} \right)^\frac{2 + \mu}{2\mu} 
\left(\frac{\rho_a}{\rho_{pj}}\right)^{\frac{\nu}{\mu}}
(g_a a_{pj} )^{-\frac{1}{\mu}}.
\label{eq:regime_transition}
\end{align}

Because of the composition and structure of primordial materials are unknown, it is difficult to definitively determine the impact regime. 
The study of the ejecta curtain from the artificial impact on asteroid 162173 Ryugu \citep{Arakawa_2020}, and the presence of ejecta near crater rims on asteroids Ryugu and 101955 Bennu \citep{Arakawa_2020,Perry_2022} infer that the strength of the materials near the surface is below 1 Pa.   The TAGSAM contact on Bennu gave a similar strength limit \citep{Walsh_2022}.  These strength limits lie below the  $\sim 25 $ Pa asteroid cohesion strength estimated from van der Waals contact forces  \citep{Scheeres_2010,sanchez_2014}.

\begin{figure}[t]
    \centering
    \includegraphics[width=7.0 cm]{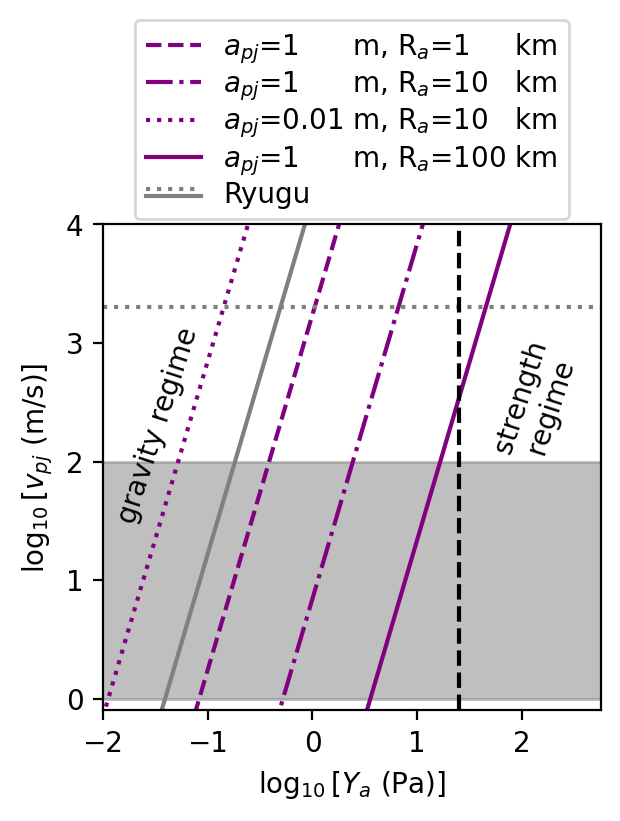}
    \caption{We plot the transition between strength and gravity regime, as determine by Equation \ref{eq:regime_transition} for$\rho_a = 235 {\rm\  kg \ m^{-3}}$ and $\pi_4=1$. The vertical dashed line is located at $Y_a = 25$ Pa, the cohesive strength from Van der Waals forces calculated by \cite{sanchez_2014}.  We highlight the rotational frame velocity ($v_{pj}$) range considered in this study as a gray band from 1 m/s to 100 m/s. We include a gray transition line for the Ryugu impact, assuming a target density of 1.79 g/cm$^3$ \citep{Miyazaki_2023}, and using the density of copper for the projectile density, along with a gray dashed line indicating the 2 km/s impact velocity of the SCI impactor \citep{Arakawa_2020}.}
    \label{fig:regime_plot}
\end{figure}

In Figure \ref{fig:regime_plot} we plot lines delineating the transition between gravity and strength regimes condition with $\rho_a = 235 {\rm\  kg \ m^{-3}}$ and $\pi_4=1$ for
different values of planetesimal and projectile radii. The gravity regime is at a higher velocity at lower surface gravity (smaller planetesimal radius), and smaller projectile radius. 
We find that for low impact velocity (between 1 and 100 m/s),
low density, planetesimal radii from 1 to 100 km, and projectiles below 1 m in radius, the impacts would mostly be in the strength regime as long as $Y_a \gtrsim 10 $ Pa.  For comparison, \cite{sanchez_2014} infer a cohesive strength of $Y_a \approx 25$ Pa for rocky rubble piles regardless of size from Van der Waals forces.  In summary, we expect the collisions considered in this study to primarily occur within the strength regime, however due to the proximity to the transition between strength and gravity regime, the crater sizes using gravity regime or strength regime do not differ greatly.

\subsection{Projectile Size} 
\label{sec:Projectile_Size}

The size distribution of projectiles is relevant for determining the impact regime, as discussed in the previous section, but also determines the size of the change in angular momentum by each impact.

Particles concentrated by the Streaming instability are expected to be within $0.04 \le {\rm St} \le 3$ \citep{Rucska_2023, Carrera_2015}. If particles experience compacted growth, are primarily porous and icy, or porous icy and experience erosion, growth rapidly occurs up until St =1, at which point inward drift begins to remove the material \citep{Krijt_2016}. Thus, we may expect an upper limit for particles with St = 1 to be present within the streaming instability in the outer disk.
Many impacts on young planetesimals are expected to be from objects that are within this range \citep{Quillen_2024}, though late states of coalescence may involve impacts from larger objects (e.g., \citealt{Nesvorny_2019}).  

Particles with $St \sim 1$ have a radius of about 1~m (based on stopping time estimates for the disk model by \citealt{Quillen_2024}). Based on stopping time arguments, we expect a minimum particle radius to be $\sim 10^{-3}$ mm \citep{Quillen_2024}. 

For projectile radii ranging from 1 cm to 1 m, assuming the fiducial values listed in Table \ref{tab:FidParams} over a range of planetesimal radii, impacts occur in a region close to the transition between strength and gravity regimes. While in the strength regime, the angular momentum transfer and mass loss is not dependent on the projectile radius. Thus, 10 collisions with a projectile of mass 100 kg, and 1 collision with a projectile of mass 1000 kg, all other parameters being equivalent, would have the same net result on ejected mass and angular momentum transfer efficiency. 

In the gravity regime, when projectile radius sets the $\pi_2$ parameter, we assume a projectile radius of $a_{pj} = 1$ m, as the largest projectile radii and subsequently projectile mass will dominate the angular momentum transfer and mass loss. Decreasing the projectile radius results in lower planetesimal strength being required for impacts to occur in the gravity regime. As such, we choose an expected particle size with the largest Stokes number to ensure we remain in the same impact regime across all planetesimal radii tested.

\section{Ejecta Distributions for Single Impacts}
\label{Sec:SingleCrater}

\begin{figure*}[htbp]
\centering
\subfloat[]{\label{fig:CraterLogVelAll}
\includegraphics[width=7.0cm]{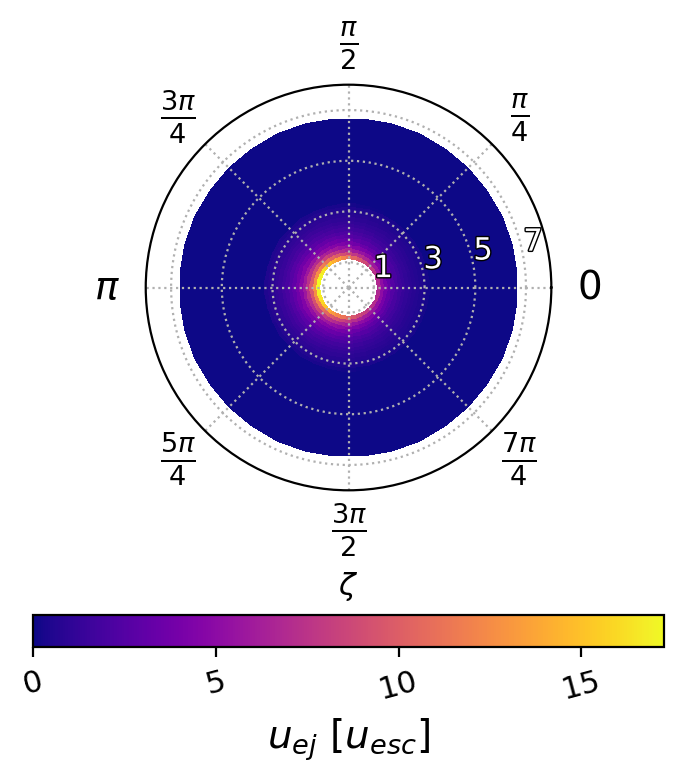}}
\subfloat[]{\label{fig:CraterLogVelEsc}\includegraphics[width=7.0cm]{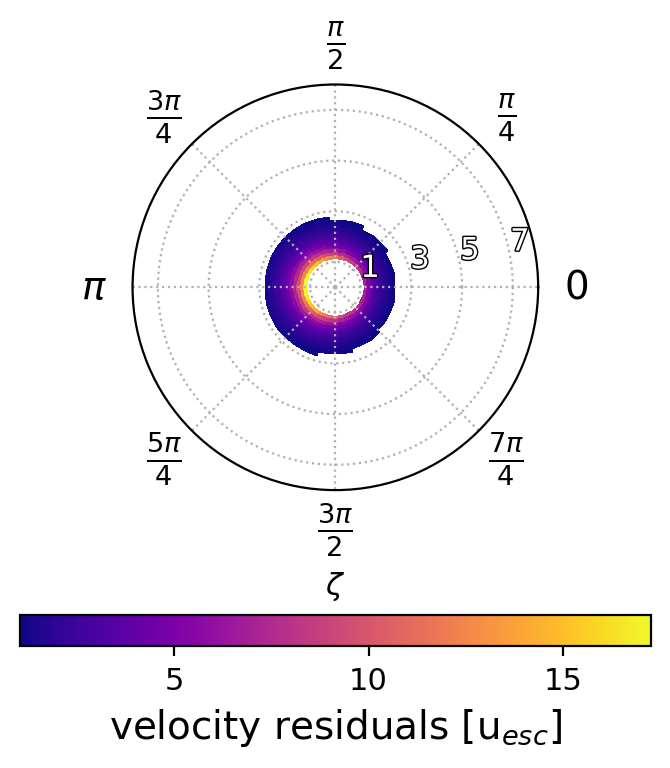}}
\caption{We show a polar plot of an impact with projectile density $\rho_{pj} = 1000$ kg m$^{-3}$ occurring at a Latitude = 0$^{\circ}$, Longitude = 0$^{\circ}$. We define Longitude = 0$^{\circ}$ as the furthest point on the headwind-facing hemisphere. (a) In the frame moving with the surface, we show ejecta velocity in color as a function of ejecta launch position and azimuthal angle.  The radius on the plot refers to distance from impact point (in units of projectile radius) and the polar angle on the plot is $\zeta$, the azimuthal angle. Due to the rotation of the planetesimal, the impact angle in the frame rotating with the surface of this collision is $\theta_I \approx 85.3^{\circ}$ despite being a head-on collision. (b) Similar to Figure \ref{fig:CraterLogVelAll}, except we only show pixels that are above the escape velocity, highlighting regions of the crater that lead to angular momentum transfer and mass loss. Despite being a head-on collision, planetesimal rotation causes an asymmetric ejecta curtain. }
\label{fig:Near-Normal}
\end{figure*} 


\begin{figure*}[!hbt]
\begin{center}
\subfloat[]{\includegraphics[width=0.3\textwidth, height=0.33\textwidth]{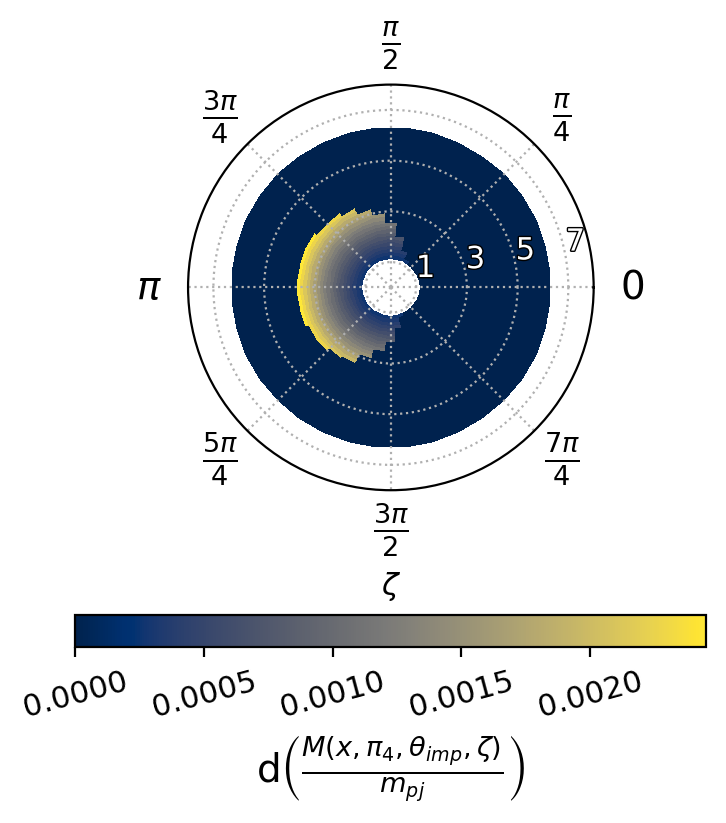}}
\subfloat[]{\includegraphics[width=0.3\textwidth,height=0.33\textwidth]{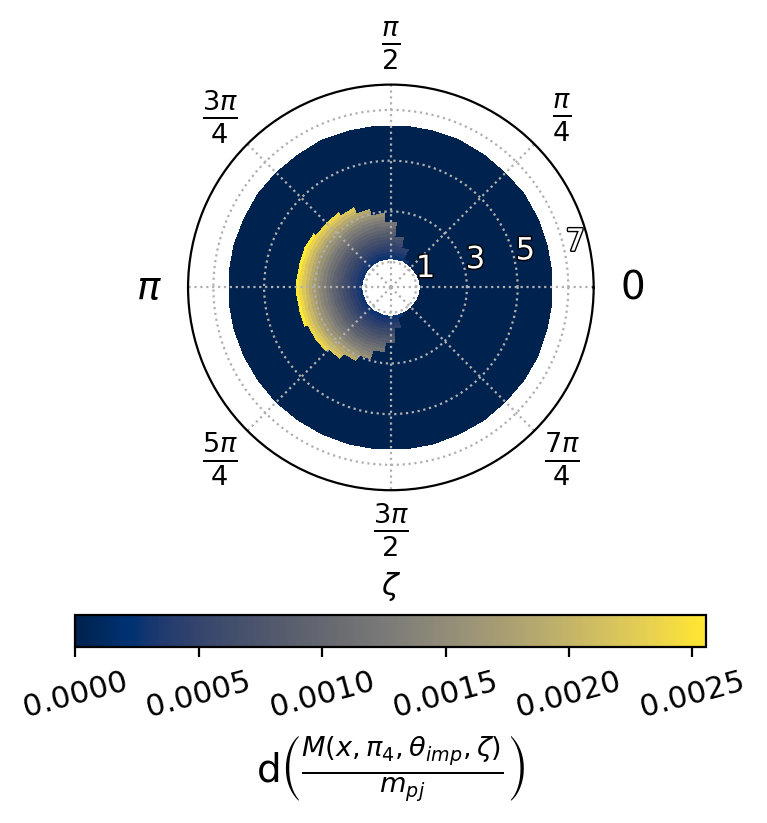}}
\\
\noindent
\subfloat[]{\includegraphics[width=0.3\textwidth, height=0.33\textwidth]{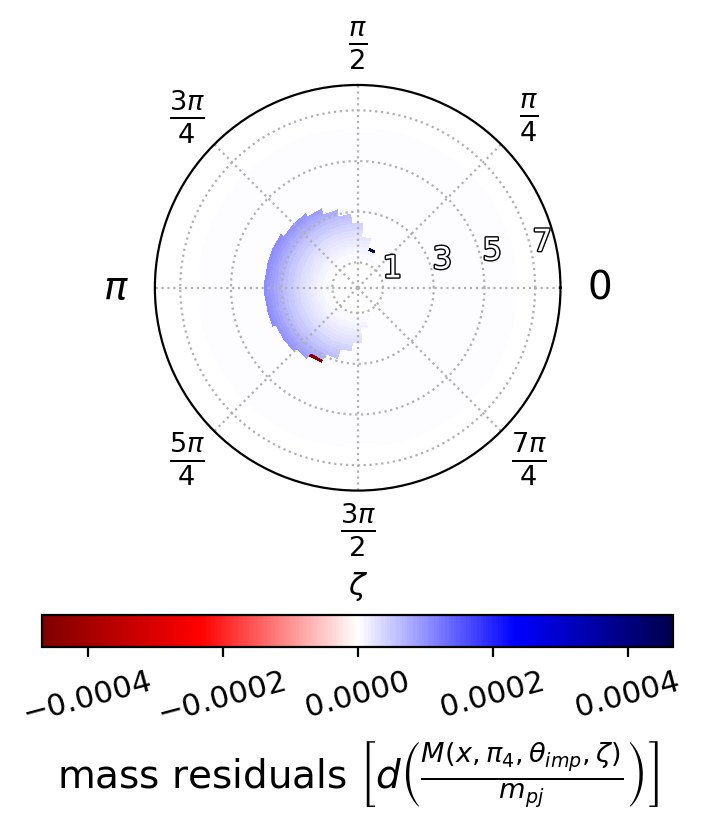}} %
\subfloat[]{\includegraphics[width=0.3\textwidth,height=0.33\textwidth]{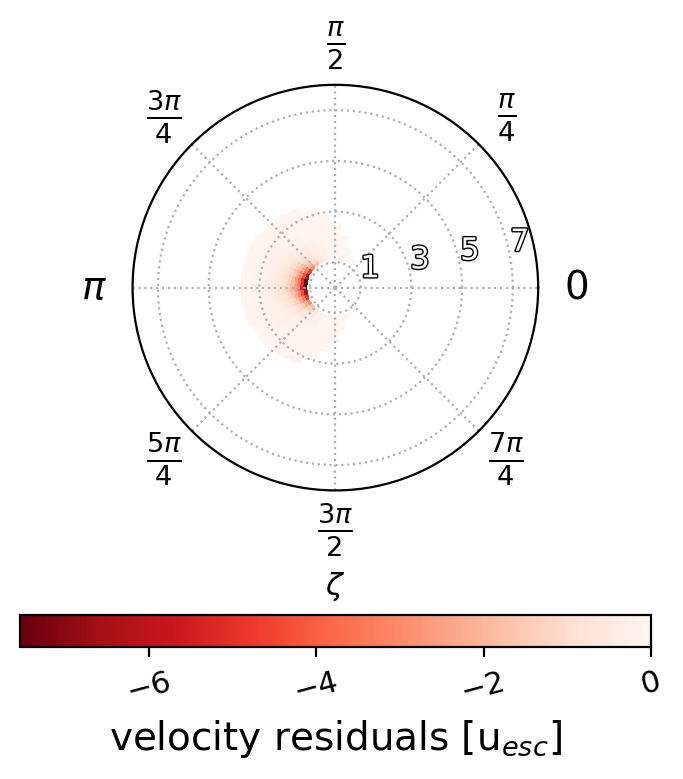}}
\end{center}
\caption[caption]{We show a polar plot of the ejecta distributions as a function of ejection position for an impact with  impact parameter $b=0.75 R_a$ and orbit angle $\alpha$ = 0, on a 20 km-radius planetesimal with and without gravitational focusing. The radius on these plots is distance from impact point and the angle, the azimuthal angle $\zeta$. (a) We show the fractions of mass ejected above escape velocity.  In this impact, gravitational focusing is neglected and the impact velocity is equal to that at large distances from the planetesimal $u_{pj} = u_{hw} = 30$ m/s. The impact occurs at surface coordinates (lat $= 0^{\circ}$, long $\sim 48.6^{\circ}$), with impact angle $\theta_{imp} \sim 48.8^{\circ}$. (b)  Similar to a) except gravitational focusing is taken into account. The  impact occurs at (lat $= 0^{\circ}$, long $\sim 47.5^{\circ}$) with $\theta_{imp} \sim 50.6^{\circ}$ with projectile velocity $u_{pj} \sim 30.86$ m s$^{-1}$. (c) Residuals of ejecta mass for the difference of (b) - (a) in units of projectile mass. (d) We show the difference in ejecta velocity in units of the escape velocity, $u_{esc}$, for the impact with gravitational focusing minus that omitting gravitational focusing. Despite the fact that the escape velocity is lower than the projectile speed distant from the planetesimal, the residuals are not small. Residuals are largest in the downrange direction. }
\label{fig:GravFocusDif}
\end{figure*} 

In Section \ref{Sec:ImpactsFrames} we explained how we estimate the total ejecta mass that escapes and the angular momentum transferred to the planetesimal from a distribution of impacts.   In this section we examine the ejecta distributions from a single impact. 
Additionally, we will explore the role of gravitational focusing. 

Unless otherwise stated, we assume a set of fiducial values for the planetesimal and projectile properties, which are listed in Table \ref{tab:FidParams}. 
Particle ejecta angles (measured from the surface normal) tend to lie between 30$^{\circ}$ - 60$^{\circ}$ with a narrow variation (10$^{\circ}$ - 20$^{\circ}$) \citep{Stoeffler_1975,Anderson_2003,Anderson_2004,Tsujido_2015,Wunnemann_2016,Luther_2018,raducan_ejecta_2022,Suo_2024}. We adopt a constant 45$^{\circ}$ ejecta angle.  

For each impact, we generate a 2 dimensional polar grid of ejecta launch positions ($x$ values) and azimuthal angles ($\zeta$ values).  Here $x$ is the radial position within the crater centered at the site of impact (see Figure \ref{fig:CraterUnits}), in units of the projectile radius, $a_{pj}$. At each point on this grid we compute an ejecta velocity in the rotating frame, an ejecta velocity vector in the inertial frame, and the ejecta mass. 
We calculate the crater radius $R_{cr,N}$ for a normal impact using Equation \ref{Eq:Rcr_str} or \ref{Eq:Rcr_grav}, depending upon whether the impact is in the strength or gravity regime. 
Following the point-source approximation, we reject locations with $x < n_1 a_{pj}$  or $x \le n_{2,N} R_{cr}$ by setting the ejecta mass and velocity at those grid points to zero. For all remaining $x$ and $\zeta$ points within our grid, we calculate the ejecta velocity and mass.  The ejecta velocity $v_{ej}$  in the rotating frame within an oblique impact crater is given by Equation \ref{Eq:ObliqueEjectaVel}, and crater radius $R_{cr}$ in equation \ref{eq:raducanRadius}.  

Using Equations \ref{Eq:vejn} and \ref{Eq:vejn_components}, we calculate the ejecta velocity components in the rotating frame at each point on our grid.
We also compute the inertial frame ejecta velocity ${\bf u}_{ej}$ (via Equations \ref{Eq:vejn}).    The ejecta velocity is compared  to the escape velocity. We assume that particles escape if $u_{ej} \ge u_{esc}$.

Figure \ref{fig:Near-Normal} shows an impact that, absent rotation, would be a normal impact at longitude (long) = 0$^{\circ}$,  latitude (lat) = 0$^{\circ}$, where we define long = 0$^{\circ}$ to be the furthest point of the planetesimal on the headwind-facing hemisphere. The influence of rotation can be seen by the fact that the ejecta curtain is not azimuthally symmetric, but rather it is oblong because it is an oblique impact, with impact angle $\theta_{imp} \approx 86.61^{\circ}$. The figure show a polar plot with angle equal to the azimuthal angle $\zeta$ and distance between the origin to the ejecta launch distance $x$ from the impact site in units of projectile radius, $a_{pj}$.  The color shows the ejecta velocity (in the rotating frame) which decreases as a function of ejecta launch distance.  

Because the impact in Figure \ref{fig:Near-Normal} is a nearly normal impact, the projectile velocity is nearly but not exactly azimuthally symmetric, as shown in Figure \ref{fig:CraterLogVelAll}. In Figure \ref{fig:CraterLogVelEsc}, the color also shows ejecta velocity in the rotating frame, but we only show points with ejecta velocity in the inertial frame that lie  above the escape velocity.  The boundary is not azimuthally symmetric. We find that the rotation is important in determining which ejecta escapes, even at low rotation velocities.
\subsection{Importance of Gravitational Focusing}

In this section
we compare ejecta distributions from a single impact assuming no gravitational deflection with one that is deflected on a hyperbolic orbit.   The projectile has density $\rho_{pj} = 1000$ kg m$^{-3}$, has impact parameter $b = 0.75 R_a$ and $\alpha = 0$ (describing the orbital tilt).  The distant headwind velocity is equal to the fiducial value of $u_{hw} = 30$ m s$^{-1}$.  The planetesimal radius in this example is $R_a = 20$ km and barely large enough that the orbital deflection could be important. Using $\rho_a = 235$ kg m$^{-3}$ and Equation \ref{Eq:EscVel}, we find that the escape velocity from the planetesimal $u_{esc} \approx 7.24$ m s$^{-1}$. 

Figure \ref{fig:GravFocusDif} compares the two ejecta distributions. Figure  \ref{fig:GravFocusDif} a and b show the ejecta mass distributions, with and without gravitational focusing,  as a function of ejection site in polar coordinates.  Consequently,  the locations and velocities of the two impacts differ.
In Figure \ref{fig:GravFocusDif}c, we subtract the ejecta mass distributions from each impact and show the mass residuals in units of $d(M(x,\pi_4,\theta_{imp}, \zeta)/m_{pj})$. As a percentage, the residual shows a maximum change of $\sim$4\% in the ejecta mass distribution as a result of gravitational focusing. In Figure \ref{fig:GravFocusDif}d, the velocity distributions are subtracted and the residual is plotted in units of escape velocity, $u_{esc}$. The residuals for ejecta velocity are large, particularly for the highest velocity material that is likely to escape the planetesimal, with the greatest residual equating to a $\sim$12\% change.

The inclusion of gravitational focusing results in a change to all of the impact parameters calculated. The impact velocity for the non-deflected case is equal to the distant headwind velocity, $u_{pj} = u_{hw}$ (30 m s$^{-1}$), whereas in the deflected case $u_{pj} \sim$ 30.86 m s$^{-1}$.
The impact locations differ slightly between the non-deflected and deflected impacts, (lat = 0$^{\circ}$, long $\sim 48.6^{\circ}$) and  (lat $ = 0^{\circ}$, long $\sim 47.5^{\circ}$), respectively. Similarly, the two impacts have different impact angles for the non-deflected and deflected impacts, ($\sim 48.8^{\circ}$ vs. $\sim 50.6^{\circ}$). The differences in impact angle (a result of different impact locations) and projectile velocity caused by gravitational focusing results in the changes in the ejecta distributions seen in Figure \ref{fig:GravFocusDif}. 

For a planetesimal with radius of $R_a = 20$ km we see the influences of gravitational focusing altering the ejecta distributions due to changes in the impact parameters. For a planetesimal with this radius (20 km), the maximum impact parameter (Equation \ref{eq:bmax}) is $b_{max} \sim 20575$ m. Thus, for a projectile with a $b/b_{max} \approx$  0.729 (Due to assuming b = 0.75 $R_a$ as was done in Figure \ref{fig:GravFocusDif}), and a planetesimal with $R_a/b_{max} \approx$  0.97, the effects of gravitational focusing are no longer negligible. 
Mild deflection of a projectile trajectory due to a low-mass planetesimal results in a noticeable change to the impact angle, velocity, and location, resulting in changes of a few percent to the ejecta distribution for the projectile.

\subsection{Caveat: Ejecta Models}

In our integrations we use an ejecta angle of 45$^\circ$; however, a distribution of angles sensitive to azimuthal angle, impact angle, and launch position would allow for more physically accurate integrations. In future, functions for ejecta scaling parameters could be fit to simulated ejecta angle distributions at a range of impact velocities (\citealt{raducan_ejecta_2022} fit 4 functions but not the ejecta angle and only simulated high velocity impacts) or to experimental data (\citealt{Anderson_2003} measured this angle  experimentally in oblique impacts but did not fit functions to them and also only considered high velocity impacts).  While \citet{Quillen_2024b} studied oblique impacts at intermediate impact velocity, they only studied the sensitivity of the integrated ejecta mass and crater radius to impact and azimuthal angle.  There is a need for more experimental characterization of oblique impacts.  

Cohesion is incorporated into a strength parameter describing the asteroid surface.  However, this description may not capture granular dynamics, particularly if projectiles are similar in size to particles on the planetesimal surface. Cohesion is often described via an energy per unit area, so it is likely to more strongly affect smaller particles as they have a large surface area relative to their mass. Experimental studies find that micron sized water-ice particles stick below a threshold velocity of 9.6 m s$^{-1}$ \citep{Gundlach_2015}.

Porosity can affect crater shape and ejecta properties \citep{Housen2003}, which we did not take into account. 
Variation due to porosity increasingly affects large craters when looking at small bodies, resulting in a reduction in ejecta mass, while for small craters the ejecta blankets are unaltered. When the product of gravitational acceleration and crater diameter, $g_aD_c, \sim 10^5$ (in cgs), for planetesimal porosity of 42\% and 70\%, the ejected crater mass begins to level out and eject 80-90\% the mass of an equivalent non-porous crater (\citealt{Housen2003}).

\section{Torque from a Distribution of Impacts} 
\label{Sec:ParamSweeps}

In the previous section we integrated the ejecta mass and velocity distributions for a single impact.  In this section we integrate a distribution of 10,000 impacts to estimate the total mass ejected and the angular momentum transferred to the planetesimal by a population of low mass impactors. 

To choose impact locations on the planetesimal surface, we generate a uniform distribution of coordinate pairs, $(x_d,y_d)$, and reject point pairs with an impact parameter $b > b_{max}$, noting that $b = \sqrt{x_d^2 + y_d^2}$. The orbit angle $\alpha = \arctan(y_d/x_d)$, and impact parameter $b$ determine the impact location,  velocity and angle, as outlined in \ref{app:Hyperbolic} and the maximum impact parameter $b_{max}$ is given in equation \ref{eq:bmax}.  

To model impacts from particles moving in a headwind, we assume that far from the planetesimal the projectiles are all unidirectional with magnitude equal to the headwind velocity, $u_{hw}$.  We take gravitational focusing into account for all planetesimal masses.  Consequently we use hyperbolic orbits to calculate the projectile impact location, velocity, and impact angle on the planetesimal surface, as described in Section \ref{sec:grav_foc}. 

The collisions are integrated to find the mass loss ratio (see Equation \ref{eqn:effiency}) and angular momentum transfer efficiency (as defined in section \ref{sec:eff}, Equations \ref{Eq:MassIntegration}, \ref{Eq:LinearMomentum} - \ref{eqn:effiency}).   We perform similar integrations but with different planetesimal and projectile properties to understand how these properties affect the angular momentum transfer efficiency and mass loss ratio.

\begin{figure}[htbp]
    \centering
    \includegraphics[width = 8cm]{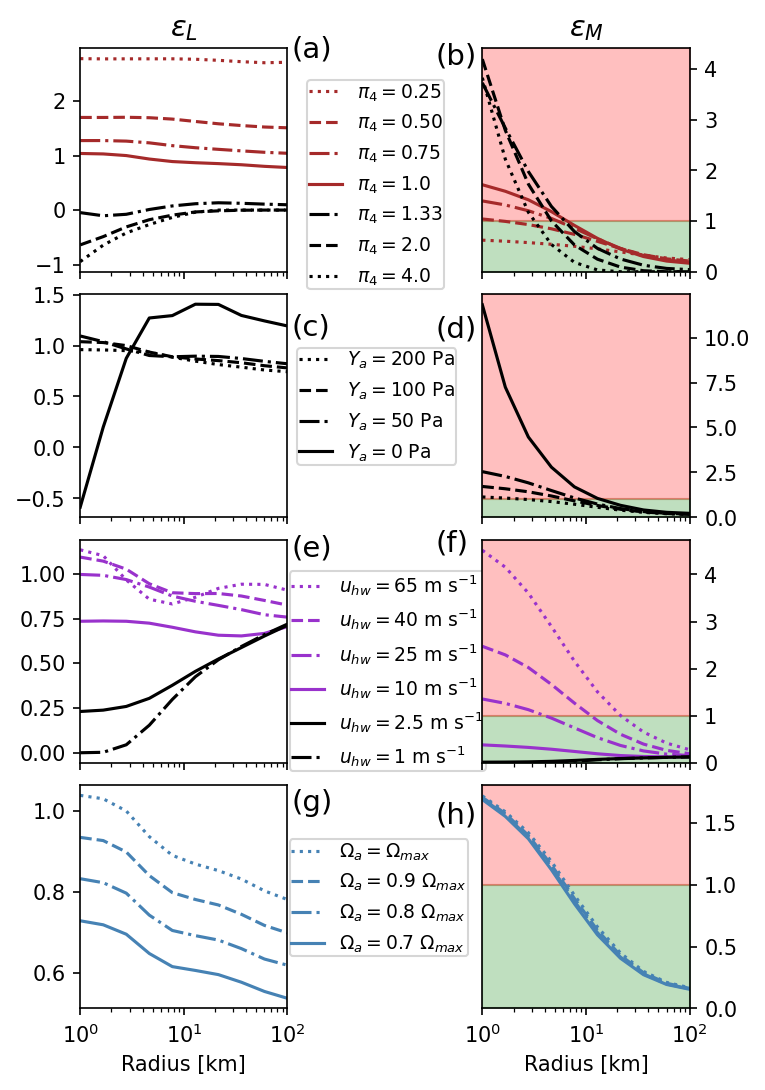}
    \caption{Mass loss ratio and angular momentum transfer efficiency for a planetesimal with the fiducial values from Table \ref{tab:FidParams} as a function of planetesimal radius on the x-axis. In each panel pair (a-b, c-d, e-f, g-h), the effect of varying a single parameter is shown. Panels (a,b) varies $\pi_4$, (b,c) varies planetesimal strength $Y_a$, (e,f) varies the distant headwind velocity $u_{hw}$, and (g,h) varies the planetesimal spin rate. The left column (subfigures a,c,e,g) show the angular momentum transfer efficiency and the right column (subfigures b, d, f, h) show the mass loss ratio. In the mass loss ratio plots b, d, f, h, we indicate regions of net accretion in green shading and net erosion in red shading. We find that the greatest influence on the angular momentum transfer are due to $\pi_4$ (the ratio of planetesimal to projectile density) and headwind velocity $u_{hw}$, while the mass loss ratio is most affected by the planetesimal strength $Y_a$.}
    \label{fig:variation}
\end{figure} 

In Figure \ref{fig:variation}, we plot the angular momentum transfer efficiency, $\epsilon_L $ and the mass loss ratio $\epsilon_M$ as a function of planetesimal radius $R_a$.  All quantities, when not specified, are taken to be those of the fiducial parameters listed in Table \ref{tab:FidParams}. In Figure \ref{fig:variation}a and \ref{fig:variation}b, we vary the projectile density $\rho_{pj}$ giving different $\pi_4$ values. These panels show how the density ratio between the projectile and planetesimal affect the mass loss ratio and angular momentum change as a function of planetesimal radius $R_a$. Figures \ref{fig:variation}c - \ref{fig:variation}h show how variations in  strength $Y_a$, headwind velocity $u_{hw}$, and spin $\Omega_a$ affect the mass loss ratio and angular momentum transfer efficiency.

The panels in the left column of Figure \ref{fig:variation} show that the angular momentum transfer efficiency is positive, confirming that the mechanism proposed by \citet{DOBROVOLSKIS1984464} operates in a low velocity regime. The panels in the right column show that, for our fiducial parameters, the total ejected mass remains in an accretionary regime for larger radii planetesimals which have higher escape velocities. At small planetesimal radii the total ejected mass is in a transitional region between erosional and accretionary dependent on the specifics of the projectile/material properties. Thus the angular momentum drain mechanism proposed by \citet{DOBROVOLSKIS1984464} also operates in an accretionary impact regime.

In Figure \ref{fig:variation}, for planetesimal radius $R_a \lesssim 5$ km, the angular momentum transfer efficiency $\epsilon_L$ is nearly independent of planetesimal radius, with Figure \ref{fig:variation}c showing a clear deviation from this overall trend, which we better describe below and in Section \ref{sec:Caveat_Oblique}. Above $R_a \sim 5$ km the angular momentum transfer efficiency tends to remain flat or slightly decrease, with the exception of low headwind velocity increasing angular momentum transfer due to gravitational focusing accelerating the projectile prior to impact.

In Figure \ref{fig:variation}a - \ref{fig:variation}b we show for a range of density ratios, with the planetesimal or projectile being 4x the density of the other, with equal densities in the middle. At small planetesimal radii, we see a trend of spinup occurring for large values of $\pi_4$. This effect is discussed in greater detail when discussing the variation in planetesimal strength and in Section \ref{sec:Caveat_Oblique}. We find that the mass loss ratio is greatest when the projectile density is less than the planetesimal for low-radii planetesimals. This is due to the mass loss ratio being the mass ejected in units of projectile mass. Decreasing the projectile mass while leaving the mass ejected unchanged will lead to a greater mass loss ratio. This is specific for low-radii planetesimals where we find a majority/all of the crater mass escapes which would occur for low-strength, small-radii planetesimals. As such, the change in projectile mass is a primary driver in the increased mass loss ratio in these cases. At larger planetesimal radii, we find that a more dense projectile results in greater a greater mass loss ratio.

In Figure \ref{fig:variation}c - \ref{fig:variation}d we find that strengthless planetesimals exhibit a much greater angular momentum transfer than planetesimals with low strength. If planetesimal strength is important in crater scaling, small variations in the planetesimal strength do not lead to large changes in the angular momentum transfer. Angular momentum drain is most effective when the cratering occurs in the gravity regime. We see a dip in the angular momentum transfer and a reversal to spin-up in the strengthless material case. This dip can begin to be seen in the low-strength case as well, showing as the small dip at $\sim 5$ km. Additionally, this dip is the same as is seen for $\pi_4 > 1$ in Figure \ref{fig:variation}a, $u_{hw} > 25$ m/s in Figure \ref{fig:variation}e, and all $\Omega_a$ in Figure \ref{fig:variation}g. This effect occurs due to highly oblique impacts in low-strength material exhibiting a butterfly crater effect, seen for oblique impacts by \citealt{Gault_1978} and \citealt{Luo_2022}, for the high-velocity ejecta. Impacts on the spin-down hemisphere see more of a butterfly effect in the escaping ejecta compared to an equivalent location impact on the spin-up hemisphere. This results in reduced angular momentum transfer in the spin-down direction, causing a decrease in net spin-down, or a complete reversal to spin-up at the lowest strength. This effect may or may not be physical, and is further discussed in Section \ref{sec:Caveat_Oblique}. At all points, especially at lower mass and radii, lower strength bodies exhibit increased mass loss because each crater sheds more material from the planetesimal when compared to one on a higher strength planetesimal. 

In Figure \ref{fig:variation}e-\ref{fig:variation}f we show how the headwind velocity affects the mass loss ratio and angular momentum transfer efficiency. Increases in the headwind velocity result in increased mass loss and angular momentum transfer. As the headwind velocity decreases, for small radii planetesimals the angular momentum transfer approaches 0, however as the planetesimal radius increases, gravitational focusing accelerates the projectiles, increasing the impact velocity. Even at low headwind velocities and planetesimal radii where gravitational focusing is not relevant, the angular momentum transfer can seem quite large. At these low headwind velocities, the ejecta velocity is smaller, thus planetesimal rotation has a greater influence in determining which ejecta escapes, as illustrated by Figure \ref{fig:EjectaFrameTransfer}. This leads to an increase in escaping ejecta on the trailing hemisphere, leading to larger spin-down. While this effect ensures that low headwind velocities can result in angular momentum drain, the decrease in ejecta velocity and leads to no ejecta escaping and thus a total shutdown of angular momentum transfer due to crater ejecta.

In Figure \ref{fig:variation}g-\ref{fig:variation}h, we show sensitivity to the planetesimal spin.  We find that the accretion rate is insensitive to the spin rate, and the angular momentum transfer efficiency decreases as a function of decreasing spin rate.  

\begin{figure}[tbp]
    \centering
    \includegraphics[width=8cm]{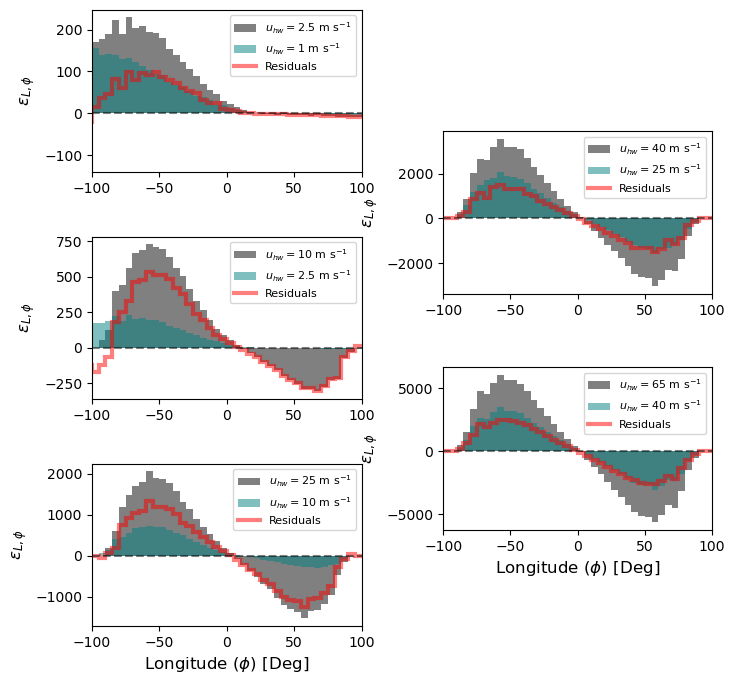}
    \caption{
    We plot a histogram of the angular momentum transfer efficiency contribution from individual longitudinal bins across the planetesimal surface. In each panel, we plot histograms at two headwind velocities, $u_{hw}$. The left side of the histograms are taller and this is why spin angular momentum is carried by the ejecta.  In each panel we show histograms at two different headwind velocities and the difference in the two histograms is shown as a red line which we label 'residual'. The darker histogram shows the higher headwind velocity. When the headwind velocity is higher, the two sides of the histogram are more symmetrical and this reduces the angular momentum lost in ejecta, even though more ejecta escapes from higher velocity impacts. This figure is consistent with the reduction in angular momentum transfer efficiency at high headwind velocities seen in Figure \ref{fig:variation}e,f. (We indicate that the Trailing Hemisphere, as shown in Figure \ref{fig:wind_on_sphere}, lies between $-100^{\circ} < \phi < 0^{\circ}$, and the Leading Hemisphere lies between $0^{\circ} < \phi < 100^{\circ}$)
    }
    \label{fig:hwvelasymm}
\end{figure}

\begin{figure}[tbp]
    \centering
    \includegraphics[width=8cm]{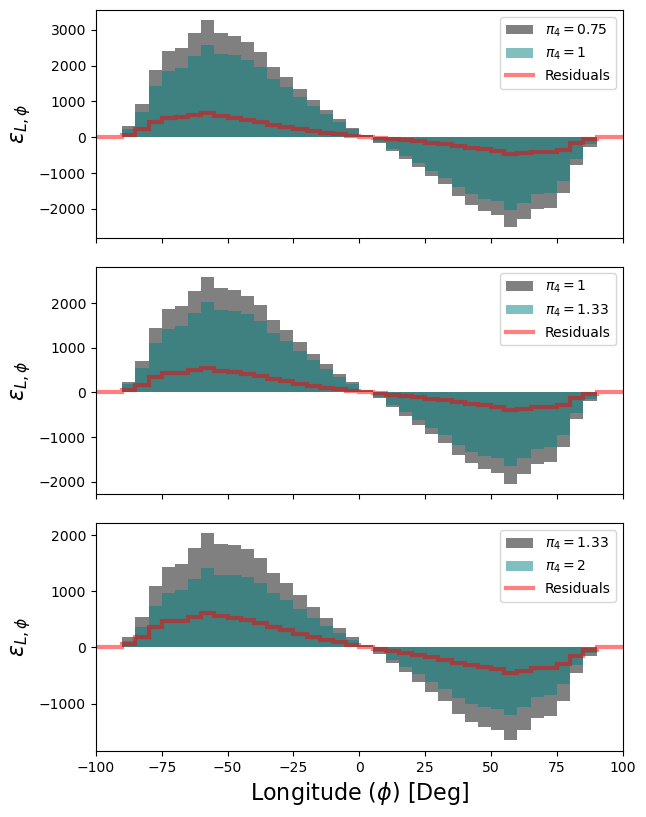}
    \caption{We plot histograms of the contribution to the angular momentum transfer efficiency from longitudinal bins across the planetesimal surface, similar to Figure \ref{fig:hwvelasymm}, except here histograms with different density ratios $\pi_4$ are shown.  When $\pi_4$ is varied, both left and right sides of the histogram vary in height, whereas in Figure \ref{fig:hwvelasymm}, only one side was affected.  }
    \label{fig:pi4symm}
\end{figure}

In Figure \ref{fig:hwvelasymm}, we show the angular momentum transfer efficiency in longitudinal bins across the planetesimal, denoted as $\epsilon_{L,\phi}$. The summation of all $\epsilon_{L,\phi}$ results in $\epsilon_L$, which we have been plotting in Figure \ref{fig:variation}. In each panel we show histograms at two different headwind velocities, and the difference in the two histograms is shown as a red line which we label `residual'. The darker histogram shows the higher headwind velocity. 
When the headwind velocity is higher, the two sides of the histogram are more symmetrical and this reduces the angular momentum lost in ejecta, even though more ejecta escapes from higher velocity impacts. This figure is consistent with the reduction in angular momentum transfer efficiency at high headwind velocities seen in Figure \ref{fig:variation}e,f.
In Figure \ref{fig:pi4symm}, is similar to Figure \ref{fig:hwvelasymm} except we show longitudinal bins for two integrations with different $\pi_4$ values. The asymmetry evident in Figure \ref{fig:pi4symm} is absent, indicating the increase in $\epsilon_L$ associated with a decrease in $\epsilon_M$ while varying $u_{hw}$. 

Among the varied parameters $\pi_4, Y_a, u_{hw}, \Omega_a$
Figure \ref{fig:variation}a shows  that 
the angular momentum transfer is most sensitive to the ratio of planetesimal to projectile density, $\pi_4$. A high density projectile colliding with a low density planetesimal will yield significantly more ejecta at high velocity and thus more angular momentum transfers than when the densities are equal. 

\subsection{A Simple Velocity Gradient}

A planetesimal embedded in a disk that is undergoing streaming instability is not an isolated object traversing through a constant headwind. A pebble cloud from which a planetesimal forms would have some inherent angular momentum of its own \citep{Nesvorny_2021, Lorek_2024}. The streaming instability generates extended concentrations of dust particles known as filaments \citep{Johansen_2007, 2017_schafer, 2015_johansen}. 
We consider a gravitationally bound pebble cloud that is within a filament which has a velocity gradient giving an angular momentum component to the projectile velocity distribution. 

We use the angular momentum of a gravitationally bound pebble cloud to estimate an upper bound for a spatial velocity gradient. If we assume that the pebble cloud extends to the Hill radius, $R_{cloud} = R_H$, where $R_H = a\big(M_{cloud}/(3M_*)\big)^\frac{1}{3}$. We take the material within the pebble cloud to be orbiting at breakup velocity, using Equation \ref{Eq:MaxRot} replacing the planetesimal density, $\rho_a$, with the modified Hill density, $\rho_{H*}$. We find that the pebble cloud angular momentum is
\begin{align}
&L_{cloud} =\frac{2}{5}M_{cloud}R_{cloud}^2\Omega_{cloud} \nonumber \\
&\sim 1.8\times10^{-5}M_{cloud}^{5/3}\ {\rm kg^{-2/3} \ m^2 \ s^{-1}} \left(\frac{a}{1{\rm AU}}\right)^{1/2}\left(\frac{M_*}{M_{\odot}}\right)^{-1/6}.
\label{eq:cloud}
\end{align}

We estimate that the filament's angular momentum is related to a deviation in the headwind velocity across the filament $\Delta u_{hw}$.
To determine the velocity gradient across the filament, we use the angular momentum density within the filament fluid flow
with the dust volume density due to the increased particle density within the streaming instability \citep{Li2021}. As such, the angular momentum density within the filament is 
\begin{equation}
l = \rho_{H*} ({\bf r}_{f}\times \Delta u_{hw}).
\end{equation}

Approximating the filament as a cylinder, we find the angular momentum within the filament to be
\begin{equation}
{\bf L}_{filament} = \rho_{H*}\int_0^{h}\int_{-R_H}^{R_H}\int_0^{\sqrt{R_H^2-y^2}} {\bf r}_{f}\times \Delta u_{hw} \ dx\ dy\ dz.
\end{equation}
Assuming the circular cross-section of the cylinder is aligned with the yz-plane, with the length of the cylinder and the direction parallel to the headwind velocity being aligned with $\hat{x}$. We assume a linear velocity gradient, $\Delta u_{hw} = \gamma y\hat{x}$, and denote ${\bf r}_{f} = y\hat{y} + z\hat{z}$. We use $\gamma$ to determine the magnitude of the linear velocity gradient. We take the length of cylinder, $h$, required to contain a total mass of $M_{cloud}$.

\begin{equation}
{\bf L}_{filament} \sim 17.6\ \gamma M_{cloud}^{5/3}\left(\frac{a}{1{\rm\ AU}}\right)^2\left(\frac{M_*}{M_{\odot}}\right)^{-2/3}\  \hat{z}.
\label{Eq:Filament}
\end{equation}
Using Equations \ref{eq:cloud} \& \ref{Eq:Filament}, we solve for the magnitude of the linear velocity gradient, $\gamma$:
\begin{equation}
\gamma \sim 1\times 10^{-6}\  {\rm s^{-1}}\left(\frac{a}{1{\rm AU}}\right)^{-3/2}\left(\frac{M_*}{M_{\odot}}\right)^{1/2}.
\label{eg:gamma_gradient}
\end{equation}

At an orbital distance of 45 AU, we approximate a velocity gradient of $\gamma \sim 3.3\times 10^{-9}\ {\rm s^{-1}}$ for a solar-mass star. The maximum change in headwind velocity, $|\Delta u_{hw}|$, experienced by a planetesimal will be at the maximum impact parameter, $b_{max}$,
\begin{align}
|\Delta u_{max}| &= \gamma b_{max} \sim 3 \times 10^{-5}\ {\rm m\ s^{-1}} \times \nonumber \\
& \left(\frac{M_{cloud}}{M_a}\right)\left(\frac{a}{45{\rm AU}}\right)^{-3/2}\left(\frac{M_*}{M_{\odot}}\right)^{-13/6} \left(\frac{R_a}{10 \ {\rm km}}\right) \times \nonumber \\ 
&\sqrt{1 + \frac{8\pi G}{3} \left(\frac{\rho_a}{235\ {\rm kg \ m^{-3}}}\right) \left(\frac{u_{hw}}{30\ {\rm m\ s^{-1}}}\right)^{-2} \left(\frac{R_a}{10\ {\rm km}}\right)^2} 
\label{eq:delta_u_max_filament}
\end{align}
which can be described as a fraction of the headwind velocity, $u_{hw}$,
\begin{equation}
    \frac{|\Delta u_{max}|}{u_{hw}} \sim 1 \times 10^{-6}.
\end{equation}

The inclusion of a velocity gradient, with lower projectile velocities on the leading hemisphere and greater velocities on the trailing hemisphere, necessitates that the angular momentum of a distribution of impacts no longer has a mean of zero. Now, the projectile distribution contains angular momentum with respect to the planetesimal, and the angular momentum transfer from the projectiles no longer averages to zero. We include the first term of Equation \ref{eqn:DeltaLa}, modifying the average change in angular momentum per impact for a distribution of particles;
\begin{align}
\langle \Delta {\bf L}_{a} \rangle = m_{pj}&\left( \bigl<
 {\bf r} \times {\bf u}_{pj} \bigl>  \ \cdot  \ 
     \hat {\boldsymbol \Omega}_a \right)  
    \hat {\boldsymbol \Omega}_a + \nonumber \\ 
     &\left( \bigl<
 {\bf r} \times ({\bf p}_{ej,esc} \cdot \hat{\bf{d}})\hat{\bf{d}} \bigl>  \ \cdot  \ 
     \hat {\boldsymbol \Omega}_a \right)  
    \hat {\boldsymbol \Omega}_a.
     \label{eqn:avg_transfer_from_pj}
\end{align}
The change in angular momentum of the planetesimal per unit projectile mass is still equal to $\langle  \Delta {\bf L}_{a}\rangle/m_{pj}$. 

\begin{figure}
    \centering
    \includegraphics[width=8cm]{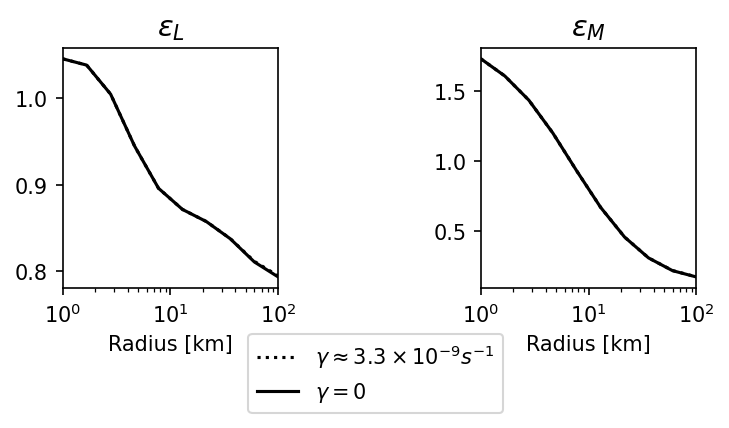}
    \caption{Angular Momentum Transfer Efficiency and Mass Loss Ratio for a planetesimal and projectile with fiducial values shown in Table \ref{tab:FidParams}. The dotted line shows the values when including a spatial velocity gradient as determined by Equation \ref{eg:gamma_gradient}. The solid line indicates the values when no velocity gradient is considered. Assuming the upper boundary for a velocity gradient remains incapable of meaningfully altering the angular momentum transfer efficiency and mass loss ratio.
    }
    \label{fig:dual_comparison}
\end{figure}

We show the maximum change in angular momentum transfer efficiency and mass loss ratio due to the inclusion of an upper bound spatial velocity gradient in Figure \ref{fig:dual_comparison}. Despite looking at the upper bound for a velocity gradient within the projectile population, no detectable changes are calculated for the angular momentum transfer efficiency, $\epsilon_L$, and the mass loss ratio, $\epsilon_M$.

Simulations of collapsing pebble clouds indicate that the angular momentum of the pebble cloud range between 0.5 - 0.01 in Hill units ($L_H = MR_H^2\sqrt{GM_{\odot}/a^3}$) \citep{Nesvorny_2021}. Converting Equation \ref{eq:cloud} to be in units of the Hill angular momentum we find $L_{cloud} \sim 2 M_{cloud}R_H^2\sqrt{GM_{\odot}/a^3}$. Thus, assuming the upper bound of a pebble cloud rotating at breakup velocity results in a value between $4\times - 200 \times$ greater than the values seen in simulations. Due to there being no meaningful change to the angular momentum transfer efficiency and mass loss ratio at our assumed upper bound, we find it unnecessary to take into account the velocity gradient in a projectile population that would be required to form the pebble cloud.

\subsection{Effective Spindown Over the Lifetime of the Collisional Era}

The spindown rate of a planetesimal is dependent on the amount of projectile mass colliding with the planetesimal, thus we need an estimate of the total mass in disk or wind particles that collide with a planetesimal in a given time frame. 
We characterize the fraction of the protostellar disk mass in particles with the ratio $f_p \equiv \rho_p/\rho_g$, where $\rho_{p}$ is the particle density in the disk (the mass per unit volume in particles), not the density of the particles themselves, and $\rho_g$ is the gas density. We use a dimensionless factor $\xi_p$ to describe the mass fraction of disk particles that can impact a planetesimal.  
Taking into account only gravitational focusing, $\xi_p = b_{max}^2 / R_a^2$.  This factor could be lower as as some particles could be swept away by gas drag. 
 
We adopt the circumstellar disk model from \citet{Quillen_2024} which has the power law form for temperature and surface density and is based on a modified version of the minimum mass solar nebula by \citet{Hayashi_1981} 
\begin{align}
    \Sigma_g &= 3430\ {\rm kg\ m}^{-2}\ g_{\Sigma}\left(\frac{a}{10 \text{AU}}\right)^{-3/2} \\
    T_g &=  170\ {\rm K} \  g_T \left(\frac{a}{1 \text{AU}}\right)^{-1/2} 
\end{align}
where $g_\Sigma$ and $g_T$
are ratios that can be used to adjust disk surface density and temperature. Following the disk model from \cite{Quillen_2024}, these ratios are equal to $1$. 
For this disk model the headwind velocity is independent of radius and $u_{hw} \approx 38$ m/s. 

Equation by \citet{Quillen_2024} estimates the mass of particles from a disk impacting the planetesimal per unit area and time at the headwind speed 
\begin{align}
    F_m & \sim  f_p \xi_p \rho_g u_{hw} \nonumber \\
    &\approx 1.28 \times 10^{-10}\  
    {\rm kg~ m}^{-2}{\rm s}^{-1} \left( \frac{f_p}{10^{-2}} \right) \nonumber\\
& \ \ \ \times \left(\frac{a}{45 \ {\rm AU} }\right)^{-\frac{11}{4}}
\frac{b_{max}^2}{R_a^2} g_\Sigma g_T^\frac{1}{2}. 
\label{eqn:Fm}
\end{align}

Assuming a planetesimal is embedded in disk undergoing streaming instability for $10^4$ years at 45 AU, with $f_p \sim 10^2$ (as streaming instability concentrates the particles), we estimate a total mass in projectiles  
\begin{align} \frac{\Delta M_a}{M_a} &= 
\frac{4\pi  R_a^2 F_m \Delta t}{M_a} \\
& \sim 0.52 \  
\left(\frac{f_p}{10^2} \right)
\left( \frac{a}{45\ {\rm AU}}\right)^{-\frac{11}{4}} g_\Sigma g_T^\frac{1}{2}  
\left(\frac{\rho_a}{235 \ {\rm kg~m}^{-3}} \right)^{-1} \nonumber \\
& \ \ \ \  \times 
\left(\frac{\Delta t}{10^4\ {\rm yr}} \right)
 \left(\frac{R_a}{10 \rm km} \right)^{-1} \left(1 + \frac{2 GM_a}{u_{hw}^2 R_a} \right) .
\end{align}

During an epoch of streaming instability, a recently formed planetesimal at 45 AU could be impacted by about half its own weight in pebbles. Depending on the specific projectile and material parameters, this mechanism could result in the removal of 30\% - 50\% of a planetesimals spin angular momentum. Angular momentum drain can occur in both accretionary regimes as well as erosional regimes, with planetesimal radii greater than 10 km always being accretionary except for $u_{hw} > 40$ m/s, and for a strengthless planetesimal, where the accretionary regime begins at slightly larger planetesimal radii. We find that for a strengthless planetesimal, while small-radii is dominated by the effects of oblique impacts (Discussed in greater detail in Section \ref{sec:Caveat_Oblique}), at large radii results in an additional $\sim$ 40\% angular momentum drain in comparison to planetesimals with some small but non-zero cohesive strength.

In this section we found that the total mass in impacts experienced by an embedded planetesimal is a fraction of its total mass. We additionally found that assuming a velocity gradient across the headwind particles does not influence the angular momentum drain nor the mass loss ratio to a meaningful degree. In our study we have found that impacts can drain a planetesimal's angular momentum, however an impact more conventionally  deflects an object. Computation of the momentum transfer efficiency would require taking into account the different directions that ejecta escape the planetesimal, during gravitational defocusing, as discussed in \ref{app:Momentum}. As the ejecta velocity is slower than the  impact velocity, gravitational defocusing could be an even stronger effect on the escaping ejecta than on the incoming projectile. 

The momentum change caused by a distribution of impacts is likely in the downwind direction, akin to a drag force.
If the planetesimal is part of binary, impact induced drag would probably aid in removing angular momentum from the binary, causing it to tighten or merge.  The large pebble mass flux present during streaming instability might facilitate binary merging.  Hence it might be interesting in a future study to  leverage ejecta scaling to compute the momentum transfer efficiency.

\subsection{Caveat: Impactor Density Distribution}
Throughout our integrations, we assumed all collisions occurred with a single projectile density $\rho_{pj}$. Since the ratio of projectile to planetesimal density, $\pi_4$, affects the integrated erosion and angular momentum transfer efficiency (See Section \ref{Sec:ParamSweeps}), the impactor density distribution would affect the total spin-down observed, due to the total projectile mass not all being of the same density. As such, evolutionary models on the size and density distribution of disk particulates (e.g. \citealt{Colmenares_2024}) could be used to expand our models.

\subsection{Caveat: Highly Oblique Impacts}
\label{sec:Caveat_Oblique}
Integrating collisions across the planetesimal surface, we expect many impacts to occur at highly oblique impact angles. The crater scaling laws we use are well tested at normal impacts \citep{Housen_2011}, and have been adjusted to account for oblique impacts down to 30 degrees \citep{raducan_ejecta_2022}. Many of the effects we see occur at these impacts with extreme obliquity. Namely, the effective spin-up for low strength planetesimals, or for large values of $\pi_4$, occur due to the ejecta velocity following a butterfly pattern at large obliquities (Impact angles below 20 - 40 degrees). As the planetesimal radii further decrease, more and more material escapes and the variations in the butterfly pattern on each hemisphere due to the effects of rotation fade away as all of the crater mass escapes. In the case for strengthless materials, this effect has not yet reversed at the lowest radii tested ($R_a = 1$ km). These results for small planetesimal radii are dominated by these highly oblique impacts, and to verify the accuracy of these results, further work is needed in low-velocity, low-strength/strengthless, highly oblique impacts.

\section{Discussion}

We discuss our work in the context of the study by \citet{Nesvorny_2021}, who studied KBO binary formation with N-body soft sphere discrete element method (PKDGRAV). 
\citet{Nesvorny_2021} found that the angular momentum of a pebble cloud prevents the rapid collapse into a single planetesimal. Instead, binary systems were more likely to form, with only a fraction of the pebble cloud being incorporated into the planetesimals. 
In this section we consider the fraction of pebble cloud angular momentum that must be removed to collapse a single planetesimal and place this value in context with the fraction of angular momentum that can be removed by escaping ejecta.

\subsection{The Angular Momentum of Planetesimals Formed Via Streaming Instability}
\label{Sec:AngMom:SI}

\citet{Nesvorny_2021} investigated 3 streaming instability simulations in which clumps formed.  They measured the angular momentum of these clumps and computed a ratio they called the mass scaled angular momentum $l  = L_c/L_H$ where $L_H$ is a Hill angular momentum and $L_c$ is
the clump angular momentum.  The Hill angular momentum $L_H  = M_c R_H^2\Omega_K$ 
where the Hill radius $R_H = a\big(M_c/(3M_*)\big)^\frac{1}{3}$ for orbital semi-major axis $a$ and $\Omega_K = \sqrt{GM_*/a^3}$
the Keplerian angular rotation rate. 
The distribution of pebble cloud $l$ values in the streaming instability simulations has a mean value of 0.1, with the full range of values between 0.01 and 0.5 (see their Figure 4). 
The ratio of the clump angular momentum to a break up value associated with a maximum spin rate 
\begin{align}
    \frac{L_c}{L_{br}} = \frac{L_c}{L_H} \frac{L_H}{L_{br}}. 
    \label{eqn:Lcbr}
\end{align}
We estimate 
$L_{br} \sim M_c R_c^2 \Omega_{break-up}$  where 
$R_c$ is a final planetesimal radius after collapse and the maximum spin value is given in equation \ref{Eq:MaxRot}. 
We find that
\begin{align}
    \frac{L_H}{L_{br}} \sim 1.5
    \left( \frac{\rho_a a^3}{M_*} \right)^\frac{1}{6}. 
\end{align}
This gives a Hill angular momentum to breakup ratio of 
\begin{align}
    \frac{L_H}{L_{br}} \sim 17
    \left( \frac{\rho_a}{1000 ~ {\rm kg~m}^{-3}}  \right)^\frac{1}{6}
    \left( \frac{M_*}{M_\odot} \right)^{-\frac{1}{6}}
    \left(\frac{a}{1~{\rm AU}} \right)^\frac{1}{2}.
\end{align}
This ratio only depends upon orbital semi-major axis and the final planetesimal density.
Inserting this into equation \ref{eqn:Lcbr} and using
the value of 0.1 for the ratio of cloud to Hill angular momentum measured in streaming instability simulations by \citet{Nesvorny_2021}, we estimate a cloud to maximum breakup angular momentum ratio of 
\begin{align} 
    \frac{L_c}{L_{br}} \sim 2
    \left( \frac{L_c/L_H }{0.1}\right) \!
    \left( \frac{\rho_a}{1000 ~ {\rm kg~m}^{-3}}  \right)^\frac{1}{6}\!\!
    \left( \frac{M_*}{M_\odot} \right)^{-\frac{1}{6}}\!\!
    \left(\frac{a}{1~{\rm AU}} \right)^\frac{1}{2}. \label{eqn:Lratio}
\end{align}

At 45 AU, $L_c / L_{br} \sim 13.4$, which is similar to the factor of $\sim$10 estimated by \citet{Nesvorny_2021} in the Kuiper Belt, implying $\sim$ 93\% of the cloud's angular momentum needs to be removed to form a single planetesimal. 

Equation \ref{eqn:Lratio} implies that the problem of angular momentum removal is worse in the outer solar system than in the inner solar system. 
In the outer solar system, due to the large fraction of angular momentum that must be lost,  the angular momentum drain mechanism investigated in this paper would account for a large fraction of the required angular momentum drain, but would be insufficient to facilitate collapse into a single planetesimal without other mechanisms.

\section{Summary}

Planetesimals embedded in protostellar disks experience impacts from disk particles. If the planetesimal is rotating, escaping ejecta can reduce its spin angular momentum, as proposed in the context of asteroids by \citet{DOBROVOLSKIS1984464}. 
In this work we integrate the momentum of ejecta from a distribution of impacts to estimate the angular momentum carried away by ejecta.  Our study differs in a number of ways from the work by \citet{DOBROVOLSKIS1984464}. Because impacts from particles in a disk would be at velocities of order 10-65 m/s rather than a few km/s,  we take into account gravitational focusing. In this study we leverage scaling laws developed for crater ejecta distributions \citep{Housen_2011} that have been modified for oblique impacts so that they are sensitive to impact and azimuthal angle \citep{raducan_ejecta_2022, Quillen_2024b}.    In contrast \cite{DOBROVOLSKIS1984464} neglected the  azimuthal angular dependence of the ejecta distributions.  Because the planetesimal has recently formed, we assume the direction from which the projectiles originate is related to the spin axis of the planetesimal and that all projectiles originate from the same direction. In contrast \citet{DOBROVOLSKIS1984464} integrated over a distribution of projectiles that come from all directions.   We consider impacts that cause accretion as well as those that cause erosion.  In contrast, at the high impact velocities considered by \citet{DOBROVOLSKIS1984464}, the impacts only cause erosion. 

We find that the angular momentum drain mechanism proposed by \citet{DOBROVOLSKIS1984464} operates on planetesimals that are embedded within a circumstellar disk and are experiencing impacts originating from a headwind.  We characterize the efficiency of the process by computing the ratio of the angular momentum lost per unit projectile mass and the planetesimal's spin angular momentum. We find that the angular momentum drain is efficient, with efficiency $\epsilon_L \sim 1$ for a wide range of parameters tested. These parameters include the projectile and planetesimal density being equal, a low-strength planetesimal with craters occurring in the strength regime, and headwind velocity being above $\sim 10$ m/s. If the projectile has greater density than the planetesimal, or if the impact occurs in the gravity regime, we find that the angular momentum drain increases, while a lower density projectile, decreased spin rate of the planetesimal, or very low headwind velocity result in decreased angular momentum drain. The expected total collisional mass is less than the mass of the planetesimal, as such we expect the total angular momentum drain to be 30 - 50\% of the initial spin angular momentum of the planetesimal. With N-body simulations \citet{Nesvorny_2021} found that 
the angular momentum of a pebble cloud prevents the collapse into a single planetesimal. Our study indicates that a pebble cloud with low initial angular momentum, with specific projectile population parameters, may collapse into single planetesimals.

We find the angular momentum transfer efficiency is most sensitive to headwind velocity and the ratio of planetesimal to projectile density. As planetesimal mass increases, the effects of gravitational focusing allow for greater angular momentum transfer from projectiles with a low headwind velocity due to accelerating the projectiles prior to impact. However, the mass loss ratio decreases as fewer particles reach the escape velocity. In the gravity regime, when projectile radius sets the $\pi_2$ parameter, we assume a projectile radius of $a_{pj} = 1$ m, as the largest projectile radii and subsequently projectile mass will dominate the angular momentum transfer and mass loss. Assuming the projectiles originate from a disk filament that might be present due to streaming instability with a velocity gradient capable of generating the computed pebble cloud, we find no meaningful reduction of the angular momentum transfer efficiency with regard to spinning down an embedded planetesimal. Future studies could take into account the angular momentum that is inherent in a population of projectiles.  

The impact related mechanism of angular momentum loss assists in removing a large portion of the excess angular momentum in a collapsing pebble cloud. However, this mechanism can not account for all of the angular momentum that needs to be removed, as such, additional sources of angular momentum drain may be necessary for the collapse of single-planetesimals. Pebble cloud collapse could lead preferentially to binary formation, as explored by \citet{Nesvorny_2019, Nesvorny_2021}. Alternatively, planetesimal formation itself could be inefficient with many pebble clouds failing to collapse, and only those that do collapse initially having low angular momentum. 

{\bf Acknowledgements:}
This material is based upon work supported by NASA grant 80NSSC21K0143.

The python code used in this study is archived at \url{https://arxiv.org/abs/2412.03533}.

\appendix
\section{The Momentum and Angular Momentum Transfer}
\label{app:Momentum}

The momentum transfer efficiency, commonly called $\beta$, is a dimensionless parameter used to characterize deflection of an asteroid by an impact (e.g., \citealt{Holsapple_2012,Feldhacker_2017,Rivkin_2021,raducan_ejecta_2022}). 
We recompute the momentum transfer parameter, taking  into account the possibility of body rotation.  We also compute the planetesimal's angular momentum change caused by the impact. 

We consider two sums, one for the initial total momentum of an impact in an inertial frame
and one for the final total momentum,  
\begin{align}
\text{Initial: \qquad }& {\bf P}_a + m_{pj} {\bf u}_{pj} + M_{ej,esc} {\bf u}_{rot}  \label{eqn:sum1}\\
\text{Final: \qquad }& {\bf P}_a + \Delta {\bf P}_a 
 + M_{ej,esc} {\bf u}_{ej,esc}. \label{eqn:sum2}
\end{align}
Here ${\bf P}_a$ is the planetesimal's initial momentum, but excluding that of the mass that will be ejected by the impact. The change of the planetesimal momentum is $\Delta {\bf P}_a$,  $m_{pj}{\bf u}_{pj}$ is the projectile momentum at the moment of impact. The velocity ${\bf u}_{rot}$ is the rotational velocity of the surface at the site of impact,  $M_{ej,esc}$ the total mass of escaping ejecta,  
and  $M_{ej,esc} {\bf u}_{rot}$ is the initial momentum of the mass in escaping ejecta but prior to impact and still on the planetesimal surface and rotating with it. 
The escaping ejecta has momentum $ M_{ej,esc}{\bf u}_{ej,esc}$ which we describe with a mean velocity ${\bf u}_{ej,esc}$ (in the inertial frame).
After impact, we assume the projectile comes to rest on the surface but if part of it is ejected, that part could
be incorporated into the escaping ejecta. 
Conservation of momentum implies that the initial and final momentum sums are equal, giving 
\begin{align}
\Delta {\bf P}_a &= m_{pj}  {\bf u}_{pj}  + M_{ej,esc}( {\bf u}_{rot} - {\bf u}_{ej,esc}) . \label{eqn:Delta_P_a}
\end{align}
The velocity difference is equivalent to a velocity in the frame rotating with the surface (via equation 
 \ref{eqn:vrotpj}). 
The momentum of escaping ejecta seen in the rotating frame 
 \begin{align} {\bf p}_{ej,esc} = 
 M_{ej,esc}( {\bf u}_{ej,esc} - {\bf u}_{rot})  = M_{ej,esc} {\bf v}_{ej,esc} \label{eqn:pesc}
 \end{align}   
 where ${\bf v}_{ej,esc}$ is the net velocity of escaping ejecta, as seen in the rotating frame. 
Thus the change in planetesimal momentum 
\begin{align}
   \Delta {\bf P}_a &= m_{pj} {\bf u}_{pj} - {\bf p}_{ej,esc}.\label{eqn:ppesc}
\end{align}
The change in planetesimal velocity  in the inertial frame, also called the deflection, would be 
\begin{align}
 \Delta {\bf V}_a = \frac{ \Delta {\bf P}_a}{M_a}
 =  \frac{ m_{pj} {\bf u}_{pj} - {\bf p}_{ej,esc}}{M_a}.
\end{align}

The momentum of escaping ejecta is often decomposed into two components, one normal to the surface and the other in the down-range or $\hat {\bf d}$ direction \citep{Rivkin_2021,raducan_ejecta_2022}.  If the planetesimal is spherical, then the  normal direction $\hat{\bf n}$ is the same in inertial and rotating frames and  the momentum transfer efficiency parameter $\beta$ is consistent with 
\begin{align}
    {\bf p}_{ej,esc} \cdot \hat {\bf n} = m_{pj} (1-\beta) (\hat {\bf n} \cdot {\bf u}_{pj}) .  \label{eqn:pej_n}
\end{align}
Neglecting rotation, we recover expressions for $\beta$ and the deflection given by \citet{Feldhacker_2017,Rivkin_2021,raducan_ejecta_2022}.

We carry out a similar calculation but for the angular momentum. We find that the change in planetesimal angular momentum is
\begin{align}
\Delta {\bf L}_a &= m_{pj} {\bf r} \times  {\bf u}_{pj}
- {\bf r} \times {\bf p}_{ej,esc} \nonumber \\
& = {\bf r} \times \Delta {\bf P}_a \label{eqn:Delta_La}
\end{align}
where $\bf r$ is the vector from the center of the planetesimal to the impact site.  
We confirm that a normal impact at longitude = 0$^{\circ}$, with projectile 
that sticks to the surface and gives no ejecta, would cause a change in angular momentum $\Delta {\bf L}_a=0$, as expected. 

Assuming that the planetesimal is experiencing many collisions, if the distribution of projectiles has no angular momentum then the $m_{pj} {\bf r} \times {\bf u}_{pj}$ term in equation \ref{eqn:Delta_La} will average to zero.
We note that in equations \ref{eqn:ppesc} and \ref{eqn:Delta_La}, the relevant escaping ejecta momentum is computed in the frame rotating with the surface at the site of impact. 

Up to this point, 
we have neglected gravitational focusing. 
Equations \ref{eqn:sum1} and \ref{eqn:sum2} describe the impact and ejecta curtain just prior to and after the impact.   The projectile velocity ${\bf u}_{pj}$ refers to velocity of the projectile relative to the planetesimal at the moment of impact (in an inertial, not rotating frame) and is not equal to the velocity of the projectile at large distance from the planetesimal. As a consequence equation \ref{eqn:pej_n} describes the momentum transfer efficiency $\beta$ only for high velocity projectiles. 

Equation \ref{eqn:Delta_La} describes the difference between the angular momentum of the planetesimal projectile just before and after the impact.  However, the gravitational force between planetesimal and projectile, which changes the relative momentum between projectile and planetesimal,  conserves angular momentum.  Thus equation \ref{eqn:Delta_La} can be used to estimate the change in angular momentum by the impact and need not take into account the trajectories of ejecta escaping the planetesimal. 

\section{The Hyperbolic Orbit Relevant for Gravitational Focusing}
\label{app:Hyperbolic}

We consider a particle approaching a planetesimal.  We neglect the gravitational force from other bodies and gas drag. The particle (or projectile)
is not gravitationally bound to the planetesimal, so it is on a hyperbolic orbit.   
We describe a hyperbolic orbit of a zero mass projectile about a planetesimal of mass $M_a$ in terms of a semi-major axis $a$ and eccentricity $e$. The semi-major axis of the hyperbolic orbit is directly related to the orbital energy and can be expressed as
\begin{align}
a = -\frac{GM_a}{u^2_{hw}}.
\end{align}
We solve for the orbital eccentricity $e$ in terms of impact parameter $b$, and the projectile relative velocity distant from the planetesimal $u_{hw}$:
\begin{align}
e^2 = 1 + \frac{b^2 u^4_{hw}}{(GM_a)^2}.
\end{align}
The distance from the center of mass of the planetesimal to the projectile is 
\begin{align}
    r(f) = \frac{a(1-e^2)}{1 + e \cos f}
\end{align}
where $f$ is the true anomaly. 

We solve for the radius where the orbit intersects the planetesimal surface, at a radius of $R_a$, giving the true anomaly of the impact site 
\begin{align}
    \cos f_{impact} = \frac{a(1-e^2)}{R_ae} - \frac{1}{e} . \label{Eq:fimp}
\end{align}
The maximum true anomaly occurs distant from the planetesimal, when
\begin{align}
    \cos f_{\infty} = - \frac{1}{e}. 
\end{align}
The angle between source wind direction (and the source of the projectile) and the site of impact, as measured from the planetesimal's center of mass, is 
\begin{align}
    \Delta f  = f_\infty - f_{impact}. \label{eqn:Delta_f}
\end{align}
This angle is used in section 
\ref{sec:grav_foc} to calculate the impact location, velocity and angle.  

\bibliographystyle{elsarticle-harv}
\bibliography{AngMomentum}


\end{document}